\documentclass[11pt,a4paper]{article}
\pdfoutput=1
\usepackage{amsfonts,amssymb,amsmath,epsfig, graphicx,yfonts,jheppub}

\title{Instability of a magnetized QGP sourced by a scalar operator}

\author{Daniel \'Avila,}
\author{Leonardo Pati\~no,}
\affiliation{Departamento de F\'isica, Facultad de Ciencias, Universidad Nacional Aut\'onoma de M\'exico, \\  A.P. 70-542, M\'exico D.F. 04510, Mexico} 

\abstract{We use the gauge/gravity correspondence to study the thermodynamics of a magnetized quark-gluon plasma in the presence of a scalar operator of dimension $\Delta=2$. We proceed by working in a five-dimensional gauged supergravity theory, where we numerically construct an asymptotically AdS$_5$ background that describes a black D3-brane in the presence of a magnetic and a scalar fields. We study the asymptotic behavior of the background and its fields close to the AdS$_5$ region to latter perform a thermodynamic analysis of the solution that includes the renormalization of the free energy associated to it. We find that because of the presence of the scalar operator, there exists a maximum intensity for the magnetic field that the plasma can hold, while for any given intensity smaller than that value, there are two states that differ in their vacuum expectation value for the scalar operator. We show that one of the two branches just mentioned is thermodynamically favored over the other.

}
  

\emailAdd{davhdz06@ciencias.unam.mx} 
\emailAdd{leopj@ciencias.unam.mx}  

\begin{document}

\maketitle
\setlength{\parskip}{3pt}

\section{Introduction and main results}

Since the first version of the gauge/gravity correspondence \cite{Maldacena:1997re} that related a conformal field theory with Anti-de Sitter space, a lot of effort has been invested to generalize this conjecture to bring the field theory closer to quantum chromodynamics (QCD). A relevant step in this direction was the inclusion of flavor \cite{Karch:2002sh} in the field theory, implemented by embedding probe branes in the dual gravitational background.

The correspondence found fertile grounds studying properties of the quark-gluon plasma (QGP) created in very high energy collisions. Part of the pursue was to find physical quantities that were robust enough to be expected to behave similarly in QCD and in the field theory accessible through the duality, where in this sense, the celebrated ratio of the entropy over the shear viscosity \cite{Policastro:2001yc,Policastro:2002se} is a prominent early example. It has been recently argued that along with the QGP, a very strong magnetic field is created for non-central collision, and furthermore, that this field can be responsible for the disagreement between certain measurements and the results expected when the magnetic field is not considered.

In \cite{DHoker:2009mmn}, a five-dimensional background was constructed so that its dual consisted of a gauge theory submerged in a constant magnetic field. This setting has been used to study a number of phenomena in a strongly interacting field theory with a background magnetic field, but if the hope is to have access to physics that include flavor, the ten-dimensional uplift of this background is necessary so that probe branes can be embedded in it. Following \cite{Cvetic:1999xp} it is simple to find the ten-dimensional solution to type IIB supergravity associated to the aforementioned background, but as we will see bellow, the particulars of the resulting geometry make it so that the usual embedding of a D7-brane is not certain to be tractable in the given parametrization.

With this in mind, we decided to use a particular way to consistently truncate ten-dimensional type IIB supergravity, knowing that in the resulting five-dimensional theory, different from the one used in \cite{DHoker:2009mmn}, it would be possible to find a family of solutions such that their uplift to 10D would be ideally suited to accommodate the embedding of a D7-brane\footnote{A different approach \cite{Filev:2007gb} can be used to introduce a magnetic field in the dual theory that includes the embedding of flavor branes from the beginning, but in this construction the U(1) field lives on the brane, and its intensity is limited by the probe approximation.}.

We were indeed able to find such a family of 5D solutions, that we present bellow, along with their 10D uplift and the embedding of probe D7-branes in it, as reported in  \cite{Avila:2019pua}. Before studying the physics involving this flavor degrees of freedom, or even to properly understand the results in \cite{Avila:2019pua}, we needed to study the thermodynamics of the backgrounds themselves, and the present work is the outcome of that analysis, that, for simplicity, we carried from the five-dimensional perspective. We would like to remark that no knowledge about the 10D uplift is necessary to reach the conclusions of the present work.

The solutions we find have a four-dimensional flat horizon, invariant under translations, and which isotropy is broken by the presence of a magnetic field that is constant in both, direction and intensity not only at the horizon, but across the whole background. The metric and the scalar field of the five-dimensional solution that extend away from the horizon depend solely on the distance from it, and in terms of this distance we will define our radial coordinate. At large radius the geometry approaches AdS$_5$ while the scalar field vanishes. All things said, these solutions represent the geometry of a black-brane, in the presence of a constant magnetic field and a radially dependent scalar field that is not minimally coupled.

The temperature associated to the horizon and the intensity of the magnetic field are two physical quantities that, from the previous description, seem appropriated to characterize each member of the family of background that we just introduced. It is also apparent that there has to be another parameter related to the scalar field, but its introduction can be done in a clearer manner in the context of the dual field theory. As we will see, the scalar field is dual to an operator ${\cal{O}}_\varphi$ of dimension $\Delta=2$, and the source of this operator will be the third parameter to fully characterize a given background.

It turns out that for any fixed temperature and value of the source of ${\cal{O}}_\varphi$, there is a maximum intensity $b_c$ of the magnetic field that the background can bare. Above $b_c$ a naked singularity is developed, while for any intensity below it, there are two solutions corresponding to two different vacuum expectation values of ${\cal{O}}_\varphi$. Our analysis permits us to determine which one of the two solutions is the thermodynamically favored, in the sense of it having a larger entropy, lower free energy, and positive specific heat. Where necessary, we will indicate this in the plots by a solid line for the preferred branch and a dash line for the unstable one.

There are two remarks we would like to make before presenting our calculations. One of them is that even if we have physical reasons to work with a vanishing source of the operator ${\cal{O}}_\varphi$, and we shall do so from section \ref{thermo} and onwards, all the results that we present here are true for any fixed finite value of it. The other thing we would like to mention is that in the process of computing the stress-energy tensor and the thermodynamic quantities that follow, there appears an arbitrary but finite contribution related to the renormalization scheme, on which some physical observables depend due to the presence of a conformal anomaly in the theory. As it happens for other backgrounds studied in the context of the gauge/gravity correspondence, in our case there is no assertive reason to fix this scheme, so we will present quantitative results for different choices, some of which have been considered in previous studies, and show that our conclusions are scheme independent.

\section{Action and solution}

As stated above, the focus of this work is the construction and analysis of the solutions of interest for \cite{Avila:2019pua}. This can be entirely done from the five-dimensional perspective, so in this section we will focus on the action and solutions in this dimensionality, and leave the introduction of flavor in the $10D$ uplift for \cite{Avila:2019pua} itself. It suffice to say for the moment that the truncation and solution ansatz that we use in this section make the $10D$ uplift particularly amiable for the embedding of D7-branes.

\subsection{$5D$ Gauged Supergravity Truncation}
The five dimensional truncation \cite{Cvetic:1999xp} that we will begin with has a matter content constituted by two independent scalar fields $\varphi_{1}$ and $\varphi_{2}$, and three independent Maxwell fields $F^{i}=dA^i$ with $i=1,2,3$, governed by the action
\begin{equation}
\begin{aligned}
S=&\frac{1}{16\pi G_{5}}\int d^{5}x\sqrt{-g}\left[R-\frac{1}{2}(\partial\varphi_{1})^{2}-\frac{1}{2}(\partial\varphi_{2})^{2}+\frac{4}{L^{2}}\sum_{i=1}^{3}X_{i}^{-1}\right]\\&-\frac{1}{16\pi G_{5}}\int\left(\frac{1}{2}\sum_{i=1}^{3}X_{i}^{-2}F^{i}\wedge \star F^{i}+F^{1}\wedge F^{2}\wedge A^{3}\right),
\end{aligned}
\label{Action}
\end{equation}
where $G_{5}$ is the five-dimensional Newton constant while
\begin{equation}
X_{i}=e^{-\frac{1}{2}\vec{a}_{i}\cdot\vec{\varphi}}, \qquad \vec{a}_{i}=(a_{i}^{(1)},a_{i}^{(2)}), \qquad \text{and} \qquad \vec{\varphi}=(\varphi_{1},\varphi_{2}).\label{truncA}
\end{equation}
Note that the choice for $A^{3}$ to appear as such in the Chern-Simons term is arbitrary, since any of the three gauge field can take its place upon performing an integration by parts. The equations of motion resulting from this action are
\begin{eqnarray}
&& \frac{1}{\sqrt{-g}}\partial_{\mu}(\sqrt{-g}g^{\mu\nu}\partial_{\nu}\varphi_{1})+\frac{2}{L^{2}}\sum_{i=1}^{3}a_{i}^{(1)}X_{i}^{-1}-\frac{1}{4}\sum_{i=1}^{3}a_{i}^{(1)}X_{i}^{-2}(F^{i})^{2}=0,
\cr
&& \frac{1}{\sqrt{-g}}\partial_{\mu}(\sqrt{-g}g^{\mu\nu}\partial_{\nu}\varphi_{2})+\frac{2}{L^{2}}\sum_{i=1}^{3}a_{i}^{(2)}X_{i}^{-1}-\frac{1}{4}\sum_{i=1}^{3}a_{i}^{(2)}X_{i}^{-2}(F^{i})^{2}=0,
\cr
&& d(X_{1}^{-2}\star F^{1})+F^{2}\wedge F^{3}=0,
\cr 
&& d(X_{2}^{-2}\star F^{2})+F^{3}\wedge F^{1}=0,
\cr
&& d(X_{3}^{-2}\star F^{3})+F^{1}\wedge F^{2}=0,
\cr
&&
\begin{aligned}
&R_{\mu\nu}-\frac{1}{2}\left(\partial_{\mu}\varphi_{1}\partial_{\nu}\varphi_{1}+\partial_{\mu}\varphi_{2}\partial_{\nu}\varphi_{2}+ \sum_{i=1}^{3}X_{i}^{-2}F^{i}_{\mu\sigma}{{F^{i}}_{\nu}}^{\sigma}\right)\\&+g_{\mu\nu}\left(\frac{4}{3L^{2}}\sum_{i=1}^{3}X_{i}^{-1}+\frac{1}{12}\sum_{i=1}^{3}X_{i}^{-2}(F^{i})^{2}\right)=0,
\end{aligned}
\end{eqnarray}
where we used the freedom discussed previously about the Chern-Simons term to write the equations of motion for the gauge fields in a more symmetric way.

As mentioned in \cite{Cvetic:1999xp}, one consistent possibility to further truncate the theory is to turn off both scalar fields and take the three Maxwell fields identical. This leads to the Einstein-Maxwell system studied in \cite{DHoker:2009mmn}, where the single $F$ was chosen to be a constant magnetic field along one of the gauge theory directions. Various physical observables have been computed using this gravitational background (see for example \cite{Arciniega:2013dqa,Arean:2016het,Avila:2018sqf,Fuini:2015hba,Endrodi:2018ikq}), but the aforementioned truncation does not permit a simple introduction of D7 flavor branes in the $10D$ geometry. The complication arises since in the procedure to introduce flavor \cite{Karch:2000gx,Karch:2002sh,Mateos:2007vn}, the D7-brane most wrap a 3-dimensional subcycle, of the 5-dimensional compact subspace $S_5$ of the background, that provides a fibration over the asymptotically AdS space and includes as a particular case a maximum subcycle of $S_5$. It can be directly seen that setting the three Maxwell fields identical to each other in the corresponding $10D$ metric on \cite{Cvetic:1999xp}, turns the identification of a 3-cycle with the desire properties into an integral problem that is not even warrantied to be solvable in this parametrization. This problem will be addressed elsewhere \cite{Uriel}.


Here we will consider a different way to further truncate (\ref{Action}) and (\ref{truncA}) given by setting
\begin{equation}
\frac{2}{\sqrt{3}}\varphi_{2}=2\varphi_{1}=\varphi, \qquad A^{1}=0, \qquad A^{2}=A^{3}=\sqrt{2}A,
\label{particular1}
\end{equation}
and keeping the vectors $\vec{a}_{i}$
\begin{equation}
\vec{a}_{1}=\left(\frac{2}{\sqrt{6}},\sqrt{2}\right), \qquad \vec{a}_{2}=\left(\frac{2}{\sqrt{6}},-\sqrt{2}\right), \qquad \vec{a}_{3}=\left(-\frac{4}{\sqrt{6}},0\right).
\label{particular2}
\end{equation}
With this choice we have
\begin{equation}
X=X_{2}=X_{3}=e^{\frac{1}{\sqrt{6}}\varphi}, \qquad X_{1}=X^{-2},
\end{equation}
and the equations of motion reduce to
\begin{eqnarray}
&& R_{\mu\nu}\!-\!\frac{1}{2}\partial_{\mu}\varphi\partial_{\nu}\varphi\!-\!2X^{-2}F_{\mu\sigma}{F_{\nu}}^{\sigma}\!+\!g_{\mu\nu}\!\left[\frac{4}{3L^{2}}\left(X^{2}\!+\!2X^{-1}\right)\!+\!\frac{1}{3}X^{-2}F_{\rho\sigma}F^{\rho\sigma}\right]=0,
\label{EOM_fondoa}\\
&& \cr
&& \frac{1}{\sqrt{-g}}\partial_{\mu}(\sqrt{-g}g^{\mu\nu}\partial_{\nu}\varphi)+\frac{4}{L^{2}}\sqrt{\frac{2}{3}}(X^{2}-X^{-1})+\sqrt{\frac{2}{3}}X^{-2}F_{\mu\nu}F^{\mu\nu}=0\label{EOM_fondob},
\end{eqnarray}
and
\begin{equation}
d(X^{-2}\star F)=0,
\label{EOM_F}
\end{equation}
along with the constrain $F\wedge F=0$, that all together can be consistently solved for $g_{\mu\nu}$, $\varphi$, and $F$.

Except for vanishing $F$ and $\varphi$, in which case the black D3-brane geometry is a solution to Eq. (\ref{EOM_fondoa}) through (\ref{EOM_F}), we have not been able to find analytic solutions to these equations of motion, so in the next subsection we will resort to a numerical approach.

We end this subsection mentioning that the consistently reduced equations of motion \eqref{EOM_fondoa}, \eqref{EOM_fondob}, and \eqref{EOM_F}, can be thought of as those coming from the effective action
\begin{equation}
S_{Eff}=\frac{1}{16\pi G_{5}}\int d^{5}x \sqrt{-g}\left[R-\frac{1}{2}(\partial\varphi)^{2}+\frac{4}{L^{2}}\left(X^{2}+2X^{-1}\right)-X^{-2}(F)^{2}\right],
\label{eff-action}
\end{equation}
that are then completed by the constrain $F\wedge F=0$, which in any case is identically satisfied by the type of solutions we study.

\subsection{Numerical solution}\label{NumSol}

For simplicity we will take $L=1$ in the following, implying that $G_{5}=\pi/2N_{c}^{2}$. We will insert in the equations of motion a similar ansatz to the one in \cite{DHoker:2009mmn}, written in a closer way to that in \cite{Arciniega:2013dqa} as
\begin{eqnarray}
&& ds^{2}=\frac{dr^{2}}{U(r)}-U(r)dt^{2}+V(r)(dx^{2}+dy^{2})+W(r)dz^{2}, 
\cr
&& F=Bdx\wedge dy,
\cr
&& \varphi=\varphi(r).
\label{metric}
\end{eqnarray}
With this choice, $F\wedge F=0$ and \eqref{EOM_F} is also automatically satisfied, while \eqref{EOM_fondoa} and \eqref{EOM_fondob} can be manipulated to give the system of differential equations
\begin{eqnarray}
&&
\begin{aligned}
&2W(r)^{2}[4B^{2}X^{-2}+V(r)(U'(r)V'(r)+U(r)V''(r))]-V(r)W(r)[2V(r)\\&\times(U'(r)W'(r)+U(r)W''(r))+U(r)V'(r)W'(r)]+U(r)V(r)^{2}W'(r)^{2}=0,
\end{aligned}
\cr
&&
\cr
&&
\begin{aligned}
&W(r)^{2}\left[V'(r)^{2}-V(r)\left(2V''(r)+V(r)\varphi'(r)^2\right)\right]-V(r)^{2}\left(W(r) W''(r)-\frac{1}{2}W'(r)^{2}\right)=0,
\end{aligned}
\cr
&& \qquad
\cr
&&
\begin{aligned}
&W(r)[-8B^{2}X^{-2}+6V(r)^{2}\left(U''(r)-\frac{8}{3}\left(X^{2}+2X^{-1}\right)\right)\\&+ 6V(r)U'(r)V'(r)]+3V(r)^{2}U'(r)W'(r)=0,
\end{aligned}\label{Equations}
\\
&& \qquad
\cr
&&
\begin{aligned}
&W(r)(\sqrt{2}B^{2}X^{-2}+V(r)^{2}\left(\frac{\sqrt{3}}{2}U'(r)\varphi'(r)+\frac{\sqrt{3}}{2}U(r)\varphi''(r)+2\sqrt{2} \left(X^{2}-X^{-1}\right)\right)\\&+\frac{\sqrt{3}}{2}U(r)V(r)\varphi'(r)V'(r))+\frac{\sqrt{3}}{4}U(r)V(r)^{2}\varphi'(r)W'(r)=0,
\end{aligned}
\cr
&& \qquad
\cr
&&
\begin{aligned}
&W(r)\left[4B^{2}X^{-2}+2V(r)U'(r)V'(r)+U(r)V'(r)^{2}-V(r)^{2}\left(U(r)\varphi'(r)^{2}+8\left(X^{2}+2X^{-1}\right)\right)\right]\\&+V(r)W'(r)\left(V(r)U'(r)+2U(r)V'(r)\right)=0,
\end{aligned}\nonumber
\end{eqnarray}
of which four are of second order, while the remaining one is of first order and plays the role of a constriction that, once satisfied at a certain radius, will hold true for any $r$.

It is important to notice that the system in \cite{DHoker:2009mmn,Arciniega:2013dqa} cannot be recovered from our current setting, since turning on any magnetic field demands a not constant, in particular not vanishing, scalar field, as can be seen for instance using our ansatz in (\ref{EOM_fondob}). Nonetheless, the system studied in \cite{Hertog:2004rz} can be recovered by setting $B=0$ and keeping $\varphi$ on, and as we will mention below, some of the conclusion for that system also apply to ours.

Since the black D3-brane geometry with $\varphi=0$ and $B=0$ is an analytic solution to our system, we will, as in \cite{Arciniega:2013dqa}, choose to write the metric functions in terms of a radial coordinate that makes them take the form
\begin{eqnarray}
&& U_{BB}(r)=\left(r+\frac{r_{h}}{2}\right)^{2}\left(1-\frac{\left(\frac{3}{2}r_{h}\right)^{4}}{\left(r+\frac{r_{h}}{2}\right)^{4}}\right),
\cr
&& V_{BB}(r)=\frac{4V_{0}}{9r_{h}^{2}}\left(r+\frac{r_{h}}{2}\right)^{2},
\cr
&& W_{BB}(r)=\frac{4}{3}\left(r+\frac{r_{h}}{2}\right)^{2},
\end{eqnarray}
with a near horizon expansion given by
\begin{eqnarray}
&& U_{BB}(r)=6r_{h}(r-r_{h})-2(r-r_{h})^{2}+\mathcal{O}(r-r_{h})^{3},
\cr
&& V_{BB}(r)=V_{0}+\frac{4V_{0}}{3r_{h}}(r-r_{h})+\frac{4V_{0}}{9r_{h}^{2}}(r-r_{h})^{2},
\cr
&& W_{BB}(r)=3r_{h}^{2}+4r_{h}(r-r_{h})+\frac{4}{3}(r-r_{h})^{2}.
\end{eqnarray}
As explained in \cite{Arean:2016het}, writing the black D3-brane solution in this manner made it possible in \cite{Arciniega:2013dqa,Arean:2016het,Martinez-y-Romero:2017awl} to work with a one parameter family of solutions that smoothly interpolates between the black brane and BTZ$\times R^2$ geometries. This interpolating family was constructed by using, as part of the numerical method that we will see below, a near horizon expansion that accommodates the behavior of both, the BTZ$\times R^2$ and black brane solutions. Even if BTZ$\times R^2$, with its vanishing $\varphi$ for a non zero $B$, is not a solution of our current theory, it will still result convenient to introduce the expansions used in \cite{Arciniega:2013dqa,Arean:2016het,Martinez-y-Romero:2017awl} and include a similar expression for the scalar, so that all together we have
\begin{eqnarray}
&& U(r)=6r_{h}(r-r_{h})+\sum_{i=2}^{\infty}U_{i}(r-r_{h})^{i},
\cr
&& V(r)=V_{0}+\sum_{i=1}^{\infty}V_{i}(r-r_{h})^{i},
\cr
&& W(r)=3r_{h}^{2}+\sum_{i=1}^{\infty}W_{i}(r-r_{h})^{i},
\cr
&& \varphi(r)=\varphi_{h}+\sum_{i=1}^{\infty}\varphi_{i}(r-r_{h})^{i},
\label{horizon_expansions}
\end{eqnarray}
and hence any member of the family of solutions has a horizon at $r_h$ with temperature given by
\begin{equation}
T= \frac{U'(r_{h})}{4\pi}=\frac{3r_{h}}{2\pi}.\label{Temp}
\end{equation}

For any solution that accepts (\ref{horizon_expansions}), the equations of motion (\ref{Equations}) are degenerated at $r_h$, so as a first step we use (\ref{horizon_expansions}) itself to solve these differential equations by a power series method near $r_h$. Following this procedure we can write all the undetermined coefficients in (\ref{horizon_expansions}), up to any desired order, in terms of the four parameters $r_h$, $B$, $V_{0}$, and $\varphi_{h}$. From here on, it will be understood that these steps have been followed and hence all the coefficients in (\ref{horizon_expansions}) are determined by the values given to $r_h$, $B$, $V_{0}$, and $\varphi_{h}$.

It would seem then, that the specific solution depends on the values of the four parameters listed in the previous paragraph, however, equations \eqref{Equations} are invariant under either simultaneous scalings of $V(r)$ and $B$ or separate scalings of $W(r)$. In consequence, non equivalent solution are obtained only for different values of the three parameters $r_h$, $B/V_{0}$, and $\varphi_{h}$, which relationship with the parameters of the dual gauge theory discussed in the introduction will be clarified below. Before proceeding any further it is worth mentioning that as a practical consistency check of our numerical codes and calculations, we constructed solutions independently varying all four values of $r_h, B, V_{0}$, and $\varphi_{h}$, and found indeed that any independent modification of $B$ and $V_{0}$ would only result in a different solution if the ratio $B/V_{0}$ changed. Without loss of generality then, $V_0$ can be set to a constant and use $B$ to control the $B/V_{0}$ parameter, reducing the number of free quantities in (\ref{horizon_expansions}) to three.

To generate a numerical solution all we need to do now is to fix some values for the three near horizon parameters in \eqref{horizon_expansions}, use these expressions to provide initial data for the metric functions and scalar field at $r=r_{h}+\epsilon$, with $\epsilon\ll r_{h}$, and then numerically integrate (\ref{Equations}) towards the boundary at $r\rightarrow \infty$.

Generically, the $r\rightarrow \infty$ behavior of the obtained metric functions is of the form
\begin{equation}
U(r)\rightarrow r^{2}, \qquad V(r)\rightarrow C_{1}r^{2}, \qquad W(r)\rightarrow C_{2}r^{2},
\end{equation}
with some constants $C_{1}$ and $C_{2}$, so the geometry approaches an scaled version of AdS$_{5}$ in what we call the near boundary region. To obtain geometries that go exactly to AdS$_{5}$ for $r\rightarrow\infty$, it is then necessary to scale the functions $V(r)$ and $W(r)$ respectively by $C_{1}$ and $C_{2}$. Given the invariance of (\ref{Equations}) that we have already mentioned, for the scaled numerical functions to still satisfy the equations of motion, $B$ has to be divided by the same factor as $V(r)$. What we end up with are numerical solutions that asymptotically approach precisely AdS$_{5}$ and have a magnetic field given by
\begin{equation}
F=b dx\wedge dy, \qquad b=\frac{B}{C_{1}}, \label{breal}
\end{equation}
which is in consequence the background magnetic field in the dual gauge theory.

\subsection{Physical parameters and maximum magnetic field}\label{PhysParam}

When following the procedure described in the last subsection it turns out that for certain combinations of $r_h$, $B/V_{0}$, and $\varphi_{h}$, the numerical solution develops a scalar field that becomes infinite and a metric function that vanishes at a finite radius greater than $r_h$. The restriction that this imposes on the set of values that $r_h$, $B/V_{0}$, and $\varphi_{h}$ can take is better understood in terms of parameters related to the behavior in the near boundary region of the solutions once they have been scaled to approach AdS$_{5}$ exactly, making then also contact with the dual gauge theory.

Of the three near horizon parameters, $r_h$ translates directly to the temperature of the horizon, and hence, through (\ref{Temp}), of the the gauge theory. The intensity $b$ of the magnetic field given in (\ref{breal}) bears information about the asymptotic behavior of the solution, since in practice $C_1=\lim_{r\rightarrow\infty}V(r)/r^2$. Finally, the behavior of the scalar field  as $r\rightarrow\infty$ is given by
\begin{equation}
\varphi\rightarrow\frac{1}{r^{2}}\left(\varphi_{0}+\psi_{0}\log{r}\right),\label{phiasy}
\end{equation}
where $\varphi_{0}$ and $\psi_{0}$ are coefficients determined by the asymtptotics of the corresponding solution. As is explained in App. \ref{AppA}, (\ref{phiasy}) implies that $\varphi$ is dual to an operator $\mathcal{O}_{\varphi}$ of dimension $\Delta=2$, and thus it saturates the BF bound \cite{Breitenlohner:1982jf}. In consequence, $\psi_{0}$ is dual to the source of the operator and $\varphi_{0}$ to its vacuum expectation value $\langle \mathcal{O}_{\varphi}\rangle$, where the precise relationship is given in \eqref{condensate}. The scaling dimension of the dual operator $\mathcal{O}_{\varphi}$ implies that it is part of a multiplet which transforms in the $\bf{20}'$ representation of $SO(6)$, meaning that it is constructed of the six adjoint scalar fields of SYM $\mathcal{N}=4$ \cite{Banks:2015aca,Banks:2016fab}.

We see that we have identified $T, b, \varphi_0$ and $\psi_0$ as four near boundary parameters, of which, according to the discussion in the previous subsection, only three can independently characterize a particular solution. From the gauge theory perspective, it makes sense to fix the temperature of the system, the intensity of the magnetic field to which it is exposed, and how much the operator ${\cal{O}}_\varphi$ is sourced, to then determine the expectation value $\langle \mathcal{O}_{\varphi}\rangle$. We shall proceed in this way.

We numerically find that for any finite value of the source at a certain temperature, there is a maximum $b_c$ for the intensity $b$ that the background can hold. For intensities beyond $b_c$, the gravitational solution develops a naked singularity, indicating that the state is unstable in the same way that it was found in 
\cite{Horowitz:2016ezu} for a critical electric field, and in 
\cite{Crisford:2018qkz} for a critical infinitesimal rotation. Both results are reflected in ours, since from the five dimensional perspective we see a critical magnetic field, while from the ten dimensional perspective this is perceived as a critical infinitesimal rotation. Just like in the two cases cited in this paragraph, we have not yet constructed a physically acceptable gravitational solution for an order parameter, $b$ in our case, higher than the critical value. The reason why this endeavor is left for future research is that, as explained in \cite{Horowitz:2016ezu}, the appearance of a naked singularity in the stationary solution indicates that the gravitational background can not be stationary for parameters higher than the critical value. Studying time dependent configurations require a separate analysis, like the one done in 
\cite{Crisford:2017zpi} to complement \cite{Horowitz:2016ezu}, and shows that the solution evolves in time to cover the singularity that started off as naked. A review of the previous results can be consulted in \cite{Crisford:2017gsb}. Given that the time dependent construction for our case is matter of future research, we limit our current calculations to states in the phase that can be investigated through the correspondence using the present gravitational configuration.

For any intensity below $b_c$, there are two solutions that differ on the value of $\langle \mathcal{O}_{\varphi}\rangle$ and, consequently, in other physical quantities associated to the state in the gauge theory that will be computed in the following sections.

\begin{figure}[ht!]
 \centering
 \includegraphics[width=0.8\textwidth]{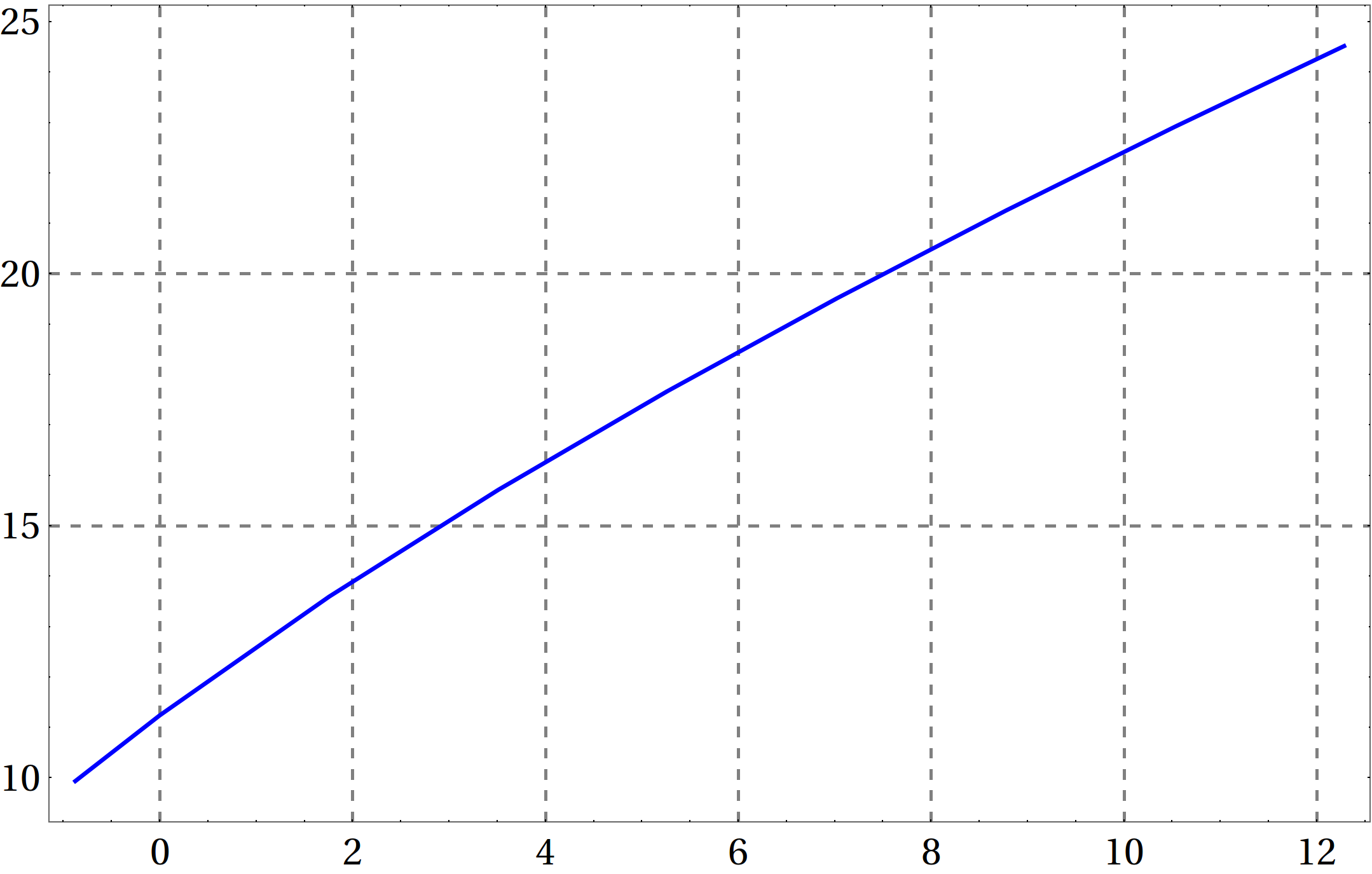}
  \put(-10,-10){\large $\psi_{0}/T^{2}$}
 \put(-390,210){\large $b_{c}/T^{2}$}
\caption{Critical magnetic field $b_{c}/T^{2}$ as a function of the source $\psi_{0}/T^{2}$ at temperature $T=3/4\pi$.}
\label{bc}
\end{figure}

Just like the critical intensity for the magnetic field, all the quantities that we investigate in this work are well defined whether the operator $\mathcal{O}_{\varphi}$ is sourced or not, but our interest is to study the theory without the deformation that a non-vanishing value of $\psi_{0}$ ads to it\footnote{Another motivation to set $\psi_{0}= 0$ is that the study 
\cite{Hertog:2004rz} of other gravitational configurations with a scalar field dual to the same operator as ours, have analytic continuations that violated cosmic censorship.}. Starting in section 3 we will set $\psi_{0}=0$, but before we dedicate the rest of the analysis exclusively to that case, we present in Fig.(\ref{bc}) the dependence of $b_c$ on the value of the source, where we see that, at least for the range that we explored numerically, it is a monotonically increasing function, with an almost linear behavior that seems to indicate that there is no bound for the critical intensity as the source is increased. Computational power did not allow us to find the value of the source for which $b_c=0$, but even if it would be interesting to know, it is not really a quantity of physical interest.

As explained in App. \ref{AppA3}, when $\psi_{0}=0$, the parameter $\varphi_0$ scales homogeneously under dilatations and in consequence, $\langle \mathcal{O}_{\varphi}\rangle/T^{2}$ only depends on the dimensionless ratio $b/T^{2}$, in terms of which all the following results will be reported. We then can, for states where $\mathcal{O}_{\varphi}$ is not sourced, fix the temperature to an arbitrary value and sweep the space of solutions using only $b$ to vary $b/T^2$. 

\begin{figure}[ht!]
 \centering
 \includegraphics[width=0.8\textwidth]{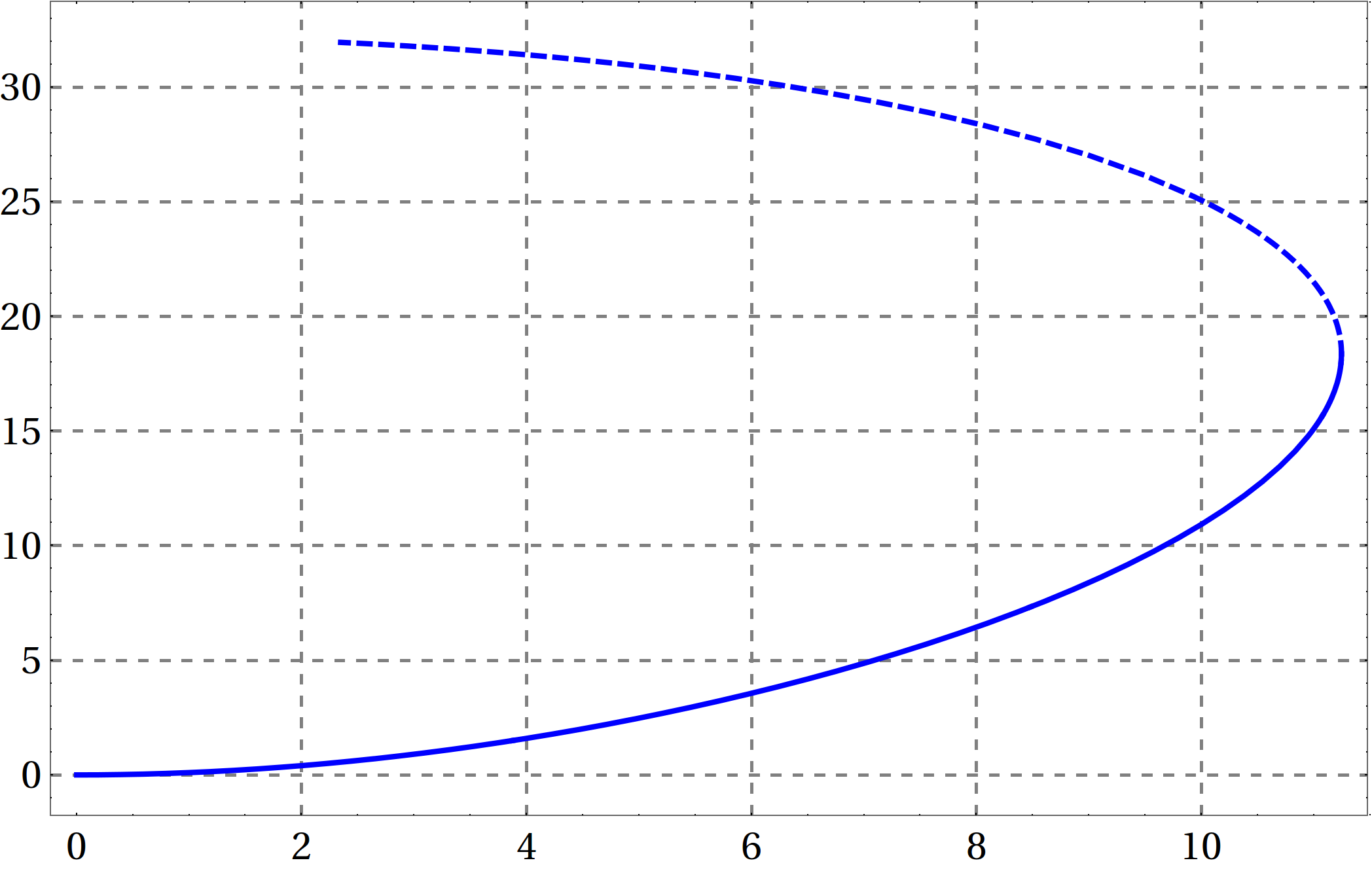}
  \put(-10,-10){\large $b/T^{2}$}
 \put(-390,210){\large $\langle \mathcal{O}_{\varphi}\rangle/T^{2}$}
\caption{Condensate $\langle \mathcal{O}_{\varphi}\rangle/T^{2}$ as a function of $b/T^{2}$ for vanishing source at temperature $T=3/4\pi$.}
\label{phiplot}
\end{figure}

Fig.(\ref{phiplot}) shows the two branches of the vacuum expectation value $\langle \mathcal{O}_{\varphi}\rangle/T^{2}$ as a function of $b/T^{2}$ up to $b_c/T^2$, with every quantity computed at vanishing source and $T=3/4\pi$.

\begin{figure}[ht!]
 \centering
 \includegraphics[width=0.8\textwidth]{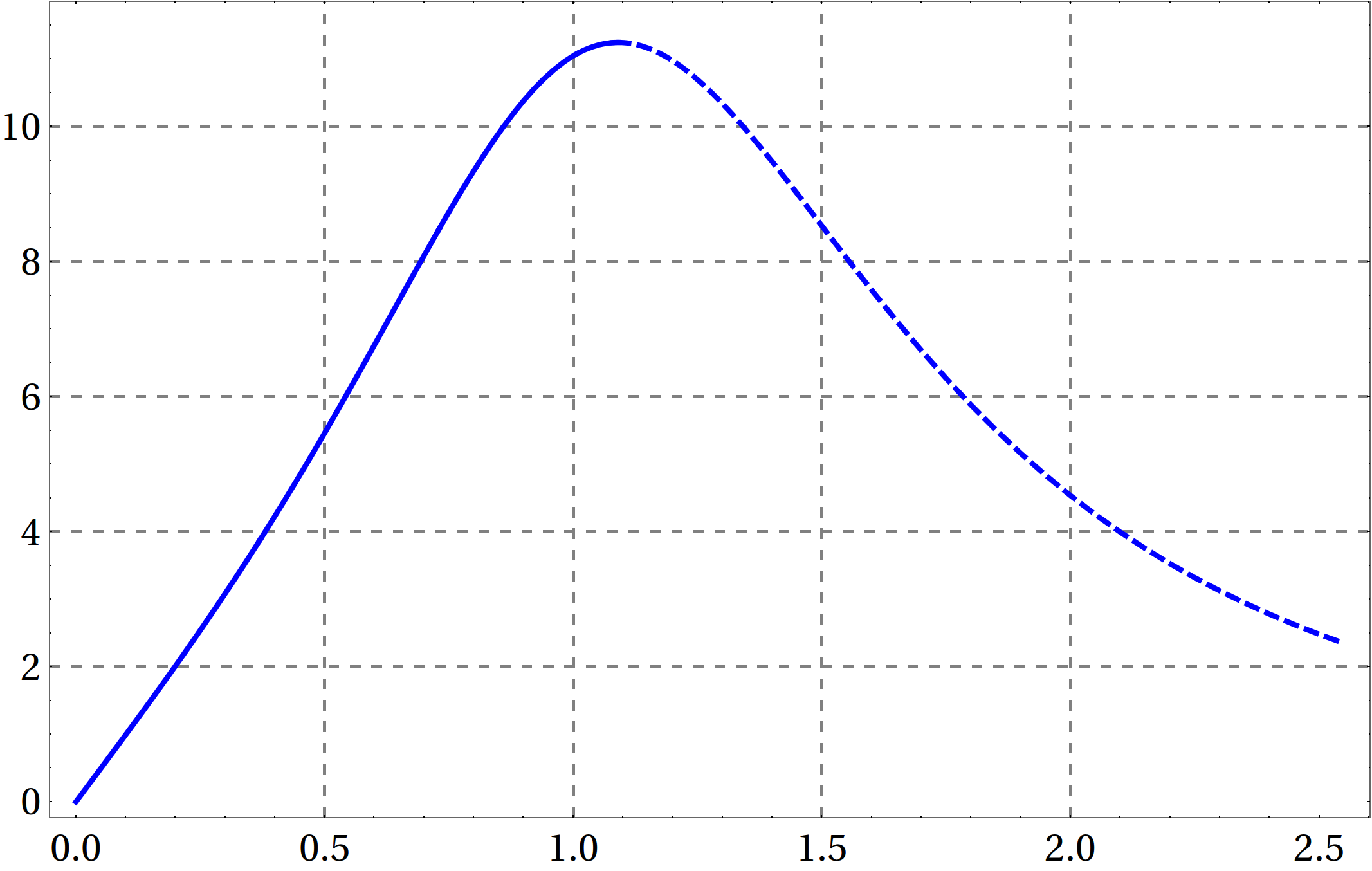}
 \put(-10,-10){\large $B$}
 \put(-370,220){\large $b/T^{2}$}
\caption{Ratio $b/T^2$ as a function of the parameter $B$ while enforcing $\psi_{0}=0$ at $T=3/4\pi$. The top of the plot defines $b_c/T^2$ at vanishing source.}
\label{bplot}
\end{figure}

Before moving on to compute other physical quantities we would like to take a moment to explain how it is that we manage to fix the source of $\mathcal{O}_{\varphi}$ to any given value, zero as a particular case, and produce plots like Fig.(\ref{phiplot}). We proceed by solving the equations of motion at given $r_h$ and $B$, for a wide range over $\varphi_{h}$ to determine the value of this last parameter that gives the desired $\psi_{0}$. To explore the full range of $b/T^2$ we keep $r_h$ fixed and repeat the procedure for as many values of $B$ as necessary to trace both branches in all our plots. It is fair to say that the two solutions for a single value of $b/T^2$ have different values of $B$ and $\varphi_{h}$, but it is important to remember that these are only auxiliary parameters in the construction of the solutions, that bare no real significance before the backgrounds have been scaled to asymptote AdS$_5$ exactly. For completeness of this constructional perspective, we should mention that, given the adjustments that are necessary to keep the source fixed as we move $B$ up from zero, the extracted constant $C_1$ in (\ref{breal}) increases too, initially growing slower than $B$, but it reaches a point at which it does so faster than this parameter, and as a consequence, $b$ begins to decrease, hence also defining $b_c$. To help visualizing the process we present Fig.(\ref{bplot}), that shows the dependence of $b/T^{2}$ on the parameter $B$ when the source is forced to stay turned off at a fixed temperature $T=3/4\pi$.

It should be clear that the procedure just described cannot lead to solutions with $b$ above $b_c$, so, if we wish to explore the instability of states violating this bound, we have to integrate the solutions from the near boundary region towards the horizon, so that the parameters that we have direct control over are $b, \varphi_0$ and $\psi_0$. The reason why we did not proceed in this way from the get go, is that finding the appropriated domain for the values of $b, \varphi_0$ and $\psi_0$ that permit gaining control over $r_h$ and impose the right conditions in the interior of the solution is harder than the way we did it, where, for instance, the position of the horizon is established at will.

What we concretely did was to extract the near boundary behavior of all the solutions close to the tip of plots like Fig.(\ref{phiplot}) to verify that if we used the asymptotic information so obtained and reverse the direction of the numerical integration, we recovered the solutions we started with. Once we had verified this, we changed the near boundary parameters slightly to get $b>b_c$, and integrate towards the horizon, confirming that in every case the solution developed a singularity before reaching it.

This completes the construction of the family of solutions, along with the appearance of a critical intensity for the magnetic field, and the double branch of solutions for intensities below it. We now proceed to compute other physical quantities related to this backgrounds and their duals in the gauge theory side.

\section{Stress-energy tensor}\label{STS}
The energy density and the pressures of the state in the gauge theory can be read from its expectation value of the stress-energy tensor, which is obtained in the dual gravitational theory as the variation of the on-shell Euclidean action with respect to the boundary metric. After performing a Wick rotation $t\rightarrow -it_{E}$ in \eqref{eff-action} we are left with
\begin{equation}
S_{E}=-\frac{1}{16\pi G_{5}}\int d^{5}x \sqrt{g}\left[R-\frac{1}{2}(\partial\varphi)^{2}+4\left(X^{2}+2X^{-1}\right)-X^{-2}(F)^{2}\right]-\frac{1}{8\pi G_{5}}\int d^{4}x \sqrt{\gamma}K,
\label{euclidean-action}
\end{equation}
where we have added the Gibbons-Hawking term, in which $\gamma$ is the determinant of the induced metric\footnote{It should properly be called a conformal structure rather than a metric, but this a common abuse of language that we will follow.} on the boundary, located at $r\rightarrow\infty$, and $K$ the trace of its extrinsic curvature. Also, it is important to remember that all the indices in \eqref{euclidean-action} are contracted with the Euclidean metric
\begin{equation}
ds^{2}=\frac{dr^{2}}{U(r)}+U(r)dt_{E}^{2}+V(r)(dx^{2}+dy^{2})+W(r)dz^{2}.
\end{equation}
We could be concerned about the fact that the constrain $F\wedge F=0$ is not derived from the action of which \eqref{euclidean-action} is the Euclidean continuation, making the appropriateness of this expression questionable. In the particular case of our solutions, the last term in \eqref{Action}, which is responsible for the constrain $F\wedge F=0$, vanishes, so the free energy can indeed be computed using \eqref{euclidean-action}.

As is commonly the case, the on-shell action suffers from near boundary divergences, that first need to be regularized by cutting the radial integral at a maximum $r$, and then subtracted by supplementing the integral with covariant boundary terms, at the hypersurface where the cut was made, to keep the result finite when the radial cut is send to infinity. This kind of procedure, to keep the on-shell action finite, is more comfortably done if the radial position is measured by the Fefferman-Graham coordinate $u$ described in App. \ref{AppA}, where we also use holographic renormalization techniques \cite{Skenderis:2002wp,Bianchi:2001kw} to show that, when written in this coordinate, the counterterms
\begin{equation}
S_{ct}=\frac{1}{16\pi G_{5}}\int d^{4}x\sqrt{\gamma}\left(6+\varphi^{2}\left(1+\frac{1}{2\log{\epsilon}}\right)+F^{ij}F_{ij}\log{\epsilon}\right),
\label{counters}
\end{equation}
are the ones we need for the full renormalized action
\begin{equation}
S=S_{E}+S_{ct},
\label{ren_action}
\end{equation}
to be finite. The integrand in (\ref{counters}) is meant to be evaluated at the radial cut-off $u=\epsilon$ and, for $S_E$ to be consistently imputed in (\ref{ren_action}), the integration over the radial direction in (\ref{euclidean-action}) has to be done from this same value $u=\epsilon$ up to the horizon at $u_h$. We remember in passing that the boundary in this coordinate is located at $u=0$.

From our results in App. \ref{AppA} it is worth mentioning that the counterterm that goes like $1/\log{\epsilon}$ appears because $\varphi$ saturates the BF bound and is only necessary when $\psi_{0}\neq 0$, while the counterterm that goes like $\log{\epsilon}$ is due to the presence of the magnetic field. The necessity for the latter is consistent with what was found in \cite{Fuini:2015hba,Endrodi:2018ikq}, where the stress-energy tensor for \cite{DHoker:2009mmn,Arciniega:2013dqa} was evaluated using similar techniques, while the presence of the former is a feature of our consistent truncation and modifies the dependence of the stress-energy tensor on the magnetic field in a non-trivial way.

We find that it is also possible to add to the action a finite term given by
\begin{equation}
S_{f}=\frac{C_{sch}}{16\pi G_{5}}\int d^{4}x\sqrt{\gamma}\left(-F^{ij}F_{ij}+\frac{\varphi^{2}}{2\log^{2}{\epsilon}}\right),
\label{finite}
\end{equation}
where the part that goes like $1/\log^{2}{\epsilon}$ is non-zero only for $\psi_{0}\neq 0$. Choosing a particular value for the free coefficient $C_{sch}$ amounts to specifying a renormalization scheme. In some circumstances it is possible to fix the scheme by demanding that the on-shell action preserves a certain symmetry. For instance, in the case of D-branes embeddings one can fix the scheme by imposing that the on-shell action vanishes for the supersymmetric embedding \cite{Karch:2005ms}. However, there are many instances where there are no such symmetries to fix the scheme, and thus it is necessary to leave $C_{sch}$ as a free parameter and study how it affects some physical observables. Such is the case in \cite{Mateos:2011ix,Mateos:2011tv}, where a thermodynamic analysis similar to the one presented here was discussed. In that work the scheme was not fixed, but was made consistent with the one used in the dual gauge theory by analyzing the chemical potential and the D7-branes sourcing the geometry. In our system there is no symmetry to fix the scheme nor source that could make the choice inconsistent with the dual gauge theory. Another possible approach is the one taken in \cite{Fuini:2015hba}, where the equivalent of $C_{sch}$ was set to $-1/4$ to simplify some expressions, and latter in \cite{Janiszewski:2015ura} the same value was adopted to eliminate, from the energy density, the explicit contribution of the electromagnetic field to the boundary stress-energy tensor. As can be seen in the appendix, $C_{sch}=-1/4$ in our case serves the same purpose as in \cite{Janiszewski:2015ura}. The hope behind this idea is to only retain the energy density associated to the plasma itself. It should be notice thought that this is not a compulsory requirement for the stress-energy tensor, and even more subtly, given how intricate the interplay between the plasma and the electromagnetic field is, removing the aforementioned explicit contribution does not guarantee that only the energy density of the plasma remains\footnote{Some details can be seen in the Appendix.}. Given the lack of an argument to fix the scheme, in the following we will carry our calculations for a number of values for $C_{sch}$, including $C_{sch}=-1/4$, making it clear in the process that the main results and conclusions obtained from our work are indeed scheme independent. 

In App. \ref{AppB} we show that the expectation value of the stress-energy tensor obtained by varying the total action
\begin{equation}
S_T=S_E+S_{ct}+S_f \label{ST}
\end{equation}
with respect to the boundary metric is given by
\begin{equation}
\begin{split}
16\pi G_{5}\langle T_{ij}\rangle=&4{g_{ij}}_{(4)}+h_{ij}(1+4C_{sch})+6C_{sch}H_{ij}\\&-{g_{ij}}_{(0)}\left({g^{kl}}_{(0)}(4{g_{kl}}_{(4)}+h_{kl})+\varphi_{(0)}(\varphi_{(0)}+\psi_{(0)}(1-\frac{2}{3}C_{sch}))\right).
\end{split}
\label{stressFGM}
\end{equation}
The right hand side of equation \eqref{stressFGM} is an expression in terms of the coefficients ${g_{ij}}_{(0)},$ ${g_{ij}}_{(4)},$  $h_{ij},$ $H_{ij},$ $\varphi_{(0)},$ and $\psi_{(0)}$, of the near boundary expansion \eqref{Fondo_FG} of the solution, done in the  FG coordinate and encoding the near boundary behavior.

Since we have better control of the numerical method when integrating out from the horizon using the coordinate $r$, we would like to find the coefficients in \eqref{stressFGM} from the solutions constructed in this way. To do so we start by expanding the equations of motion \eqref{Equations} around $r=\infty$ and solving them order by order in $1/r$, obtaining, after imposing exact AdS$_5$ asymptotics and  vanishing source $\psi_0$, the expressions
\begin{eqnarray}
&& U(r)=r^{2}+U_{1}r+\frac{U_{1}^{2}}{4}+\frac{1}{r^{2}}\left(U_{4}-\frac{2}{3}b^{2}\log{r}\right)+\mathcal{O}\left(\frac{1}{r^{4}}\right),
\cr
&& V(r)=r^{2}+U_{1}r+\frac{U_{1}^{2}}{4}+\frac{1}{r^{2}}\left(-\frac{1}{2}W_{4}-\frac{1}{6}\varphi_{0}^{2}+\frac{1}{3}b^{2}\log{r}\right)+\mathcal{O}\left(\frac{1}{r^{4}}\right),
\cr
&& W(r)=r^{2}+U_{1}r+\frac{U_{1}^{2}}{4}+\frac{1}{r^{2}}\left(W_{4}-\frac{2}{3}b^{2}\log{r}\right)+\mathcal{O}\left(\frac{1}{r^{4}}\right),
\cr
&& \varphi(r)=\frac{\varphi_{0}}{r^{2}}-\frac{U_{1}\varphi_{0}}{r^{3}}+\frac{1}{12r^{4}}\left(-2\sqrt{6}b^{2}+\varphi_{0}(9U_{1}^{2}-\sqrt{6}\varphi_{0})\right)+\mathcal{O}\left(\frac{1}{r^{5}}\right),
\label{r_expansions}
\end{eqnarray}
where $U_{1}$, $U_{4}$, $W_{4}$ and $\varphi_{0}$ are the coefficients, not determined by the equations of motion, that have to be read from the numerical solution associated to each particular value of $b$ and $T$, making them functions of this physical parameters. Once these coefficients have been extracted, we can use the relationships
\begin{eqnarray}
&& u(r)=\frac{1}{r}-\frac{U_{1}}{2r^{2}}+\frac{U_{1}^{2}}{4r^{3}}-\frac{U_{1}^{3}}{8r^{4}}+\frac{1}{r^{5}}\left(\frac{1}{48}(b^{2}+3U_{1}^{4}-6U_{4})+\frac{1}{12}b^{2}\log{r}\right)+\mathcal{O}\left(\frac{1}{r^{6}}\right),
\cr
&& r(u)=\frac{1}{u}-\frac{U_{1}}{2}+u^{3}\left(\frac{1}{48}(b^{2}-6U_{4})-\frac{1}{12}b^{2}\log{u}\right)+\mathcal{O}(u^{5}),\label{u2r}
\end{eqnarray}
that $r$ and $u$ hold close to the boundary, to eliminate $r$ in \eqref{r_expansions} in favor of $u$, obtaining the expansions in terms of $u$ of the metric functions and scalar field given by
\begin{eqnarray}
&& U(u)=\frac{1}{u^{2}}+u^{2}\left(\frac{1}{24}b^{2}+\frac{3}{4}U_{4}+\frac{1}{2}b^{2}\log{u}\right)+\mathcal{O}(u^{4}),
\cr
&& V(u)=\frac{1}{u^{2}}+u^{2}\left(\frac{1}{24}b^{2}-\frac{1}{4}U_{4}-\frac{1}{2}W_{4}-\frac{1}{6}\varphi_{0}^{2}-\frac{1}{2}b^{2}\log{u}\right)+\mathcal{O}(u^{4}),
\cr
&& W(u)=\frac{1}{u^{2}}+u^{2}\left(\frac{1}{24}b^{2}-\frac{1}{4}U_{4}+W_{4}+\frac{1}{2}b^{2}\log{u}\right)+\mathcal{O}(u^{4}),
\cr
&& \varphi(u)=u^{2}\varphi_{0}+u^{4}\left(-\frac{b^{2}}{\sqrt{6}}-\frac{\varphi_{0}^{2}}{2\sqrt{6}}\right)+\mathcal{O}(u^{6}).
\label{FG_num}
\end{eqnarray}

The expansion \eqref{FG_num} in terms of the coefficients $U_{1}$, $U_{4}$, $W_{4}$ and $\varphi_{0}$, has to be the same as the one given by \eqref{Fondo_FG} in terms of the coefficients ${g_{ij}}_{(0)},$ ${g_{ij}}_{(4)},$  $h_{ij},$ $H_{ij},$ $\varphi_{(0)},$ and $\psi_{(0)}$, so we can solve for the latter in terms of the former and evaluate \eqref{stressFGM} to extract the energy and pressures
\begin{equation}
\langle T_{ij} \rangle=\text{diag}(E,P^{\perp},P^{\perp},P^{\parallel}),\label{diagT}
\end{equation}
as functions of $b$ and $T$, resulting in
\begin{eqnarray}
&& E=\frac{N_{c}^{2}}{8\pi^{2}}\left(-3U_{4}-\frac{1}{3}\varphi_{0}^{2}-2C_{sch}b^{2}\right),
\cr
&& P^{\perp}=\frac{N_{c}^{2}}{8\pi^{2}}\left(-U_{4}-2W_{4}-\frac{1}{3}\varphi_{0}^{2}-b^{2}(1+2C_{sch})\right),
\cr
&& P^{\parallel}=\frac{N_{c}^{2}}{8\pi^{2}}\left(-U_{4}+4W_{4}+\frac{1}{3}\varphi_{0}^{2}+2C_{sch}b^{2}\right),
\label{Stress}
\end{eqnarray}
where $P^{\perp}$ and $P^{\parallel}$ are respectively the pressures along directions perpendicular and parallel to the magnetic field. The schematic form of these expressions reduces to the ones reported in \cite{Fuini:2015hba,Endrodi:2018ikq} when taking $\varphi_{0}=0$, however, we stress that this cannot be done in consistency with the equations of motion unless we also demand $b=0$, confirming again that the solutions studied in \cite{DHoker:2009mmn,Arciniega:2013dqa} cannot be recovered from our setting.

Something important that emerges from the previous results is that there is a conformal anomaly in our theory, revealed by a non vanishing trace of the expectation value of the stress-energy tensor. Using \eqref{diagT} and  \eqref{Stress}, this trace can be computed to be
\begin{equation}
{\langle T^i}_{i}\rangle=-\left( \frac{N_{c}\,b}{2\pi}\right) ^{2},
\end{equation}
where we see that, from our physical parameters, the anomaly only depends quadratically on $b$, and disappears for vanishing magnetic field, resembling the massless QED result for the trace anomaly which is given by 
\begin{equation}
\Theta^i_i=\frac{\beta(e)}{2e^3}F^2,
\end{equation}
where $e$ is the electric charge. This result coincide with \cite{Fuini:2015hba,Endrodi:2018ikq}, where the trace anomaly of the theory was also found to be quadratic on the intensity of the magnetic field. We note here that our result would be modified if the source for the scalar operator was non-zero, as can be seen from \eqref{traceFG}.

Below we show the numerical results for the components of the expectation value of the stress-energy tensor for four different renormalization schemes, given by  $C_{sch}=\{ -5, -1/4, 0, 5\}$, normalized with respect to its values for $b=0$ and $\varphi=0$
\begin{equation}
E_{0}=\frac{3\pi^{2}N_{c}^{2}}{8}T^{4}, \qquad P_{0}=\frac{\pi^{2}N_{c}^{2}}{8}T^{4}.
\end{equation}
\begin{figure}[ht!]
 \centering
 \includegraphics[width=0.8\textwidth]{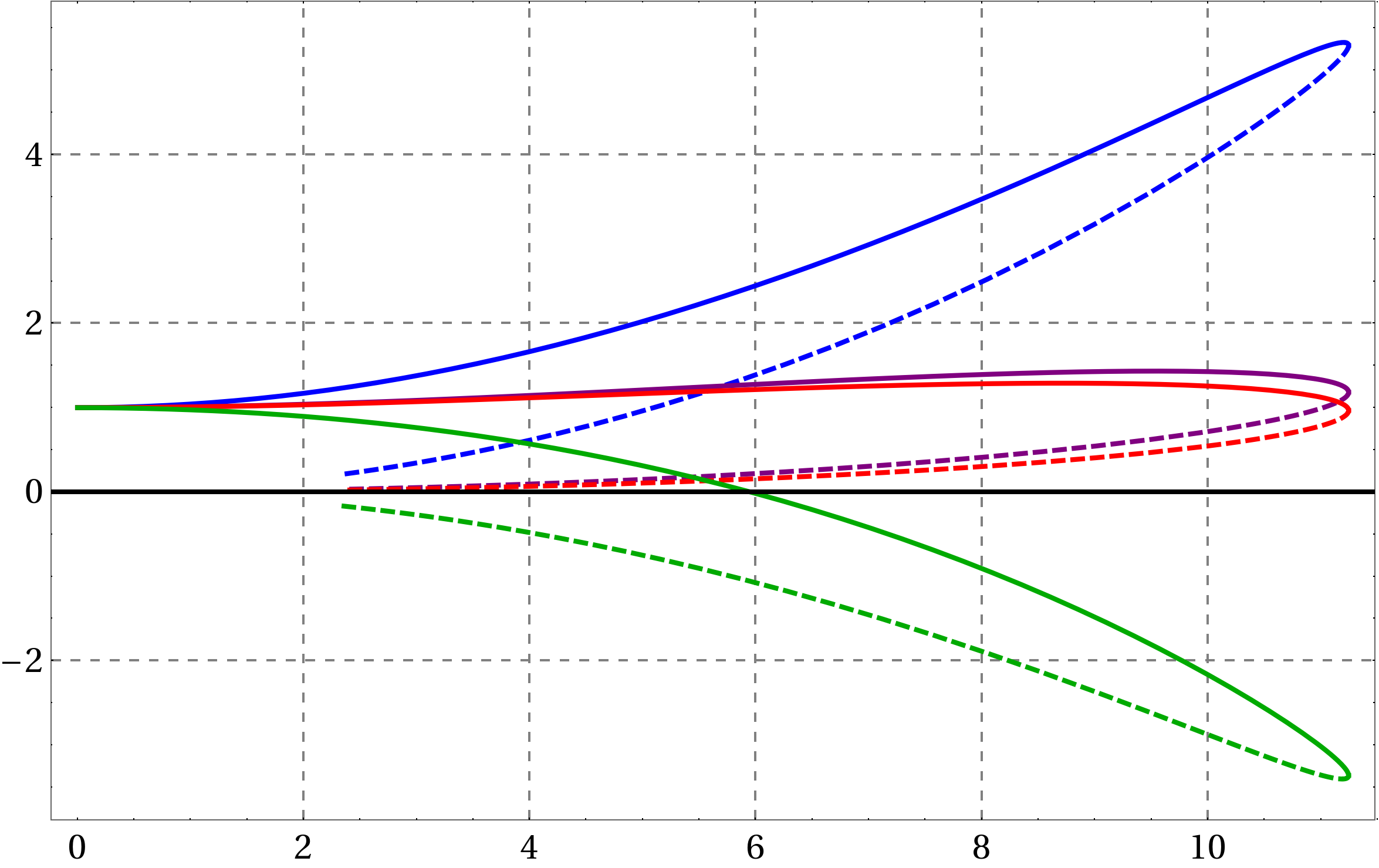}
 \put(-10,-10){\large $b/T^{2}$}
 \put(-370,220){\large $E/E_{0}$}
\caption{Energy density $E$ as a function of $b/T^{2}$, normalized with respect to its value $E_{0}$ at $b=0$ and $\varphi=0$. Each curve corresponds to a different renormalization scheme given by  $C_{sch}=-5$ (blue), $C_{sch}=-1/4$ (purple), $C_{sch}=0$ (red), and $C_{sch}=5$ (green).}
\label{Energy}
\end{figure}
\begin{figure}[ht!]
 \centering
 \includegraphics[width=0.8\textwidth]{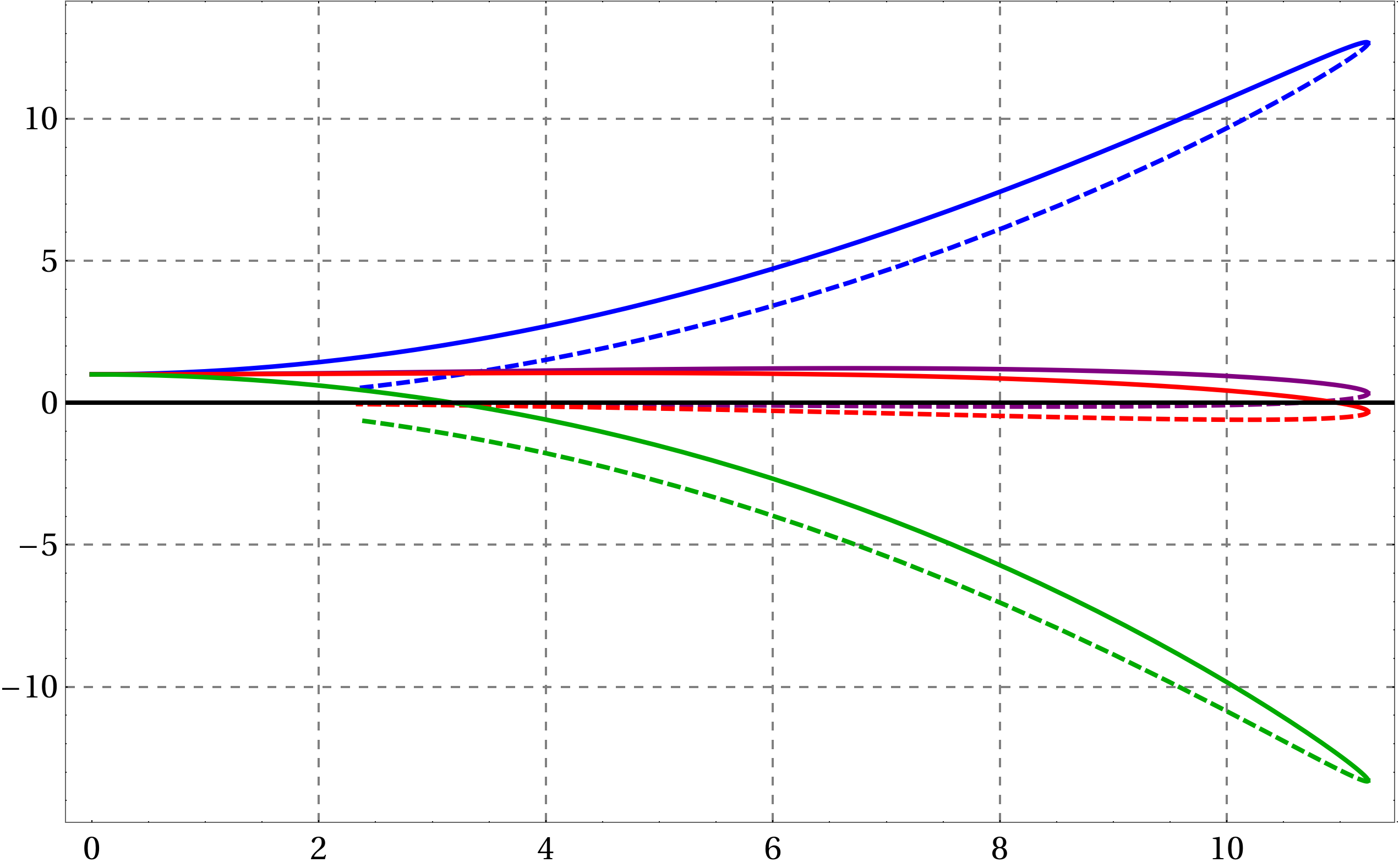}
 \put(-10,-10){\large $b/T^{2}$}
 \put(-370,220){\large $P^{\perp}/P_{0}$}
\caption{Pressure $P^{\perp}$ as a function of $b/T^{2}$, normalized with respect to its value $P_{0}$ at $b=0$ and $\varphi=0$. Each curve corresponds to a different renormalization scheme given by  $C_{sch}=-5$ (blue), $C_{sch}=-1/4$ (purple), $C_{sch}=0$ (red), and $C_{sch}=5$ (green).}
\label{PresurePerp}
\end{figure}
\begin{figure}[ht!]
 \centering
 \includegraphics[width=0.8\textwidth]{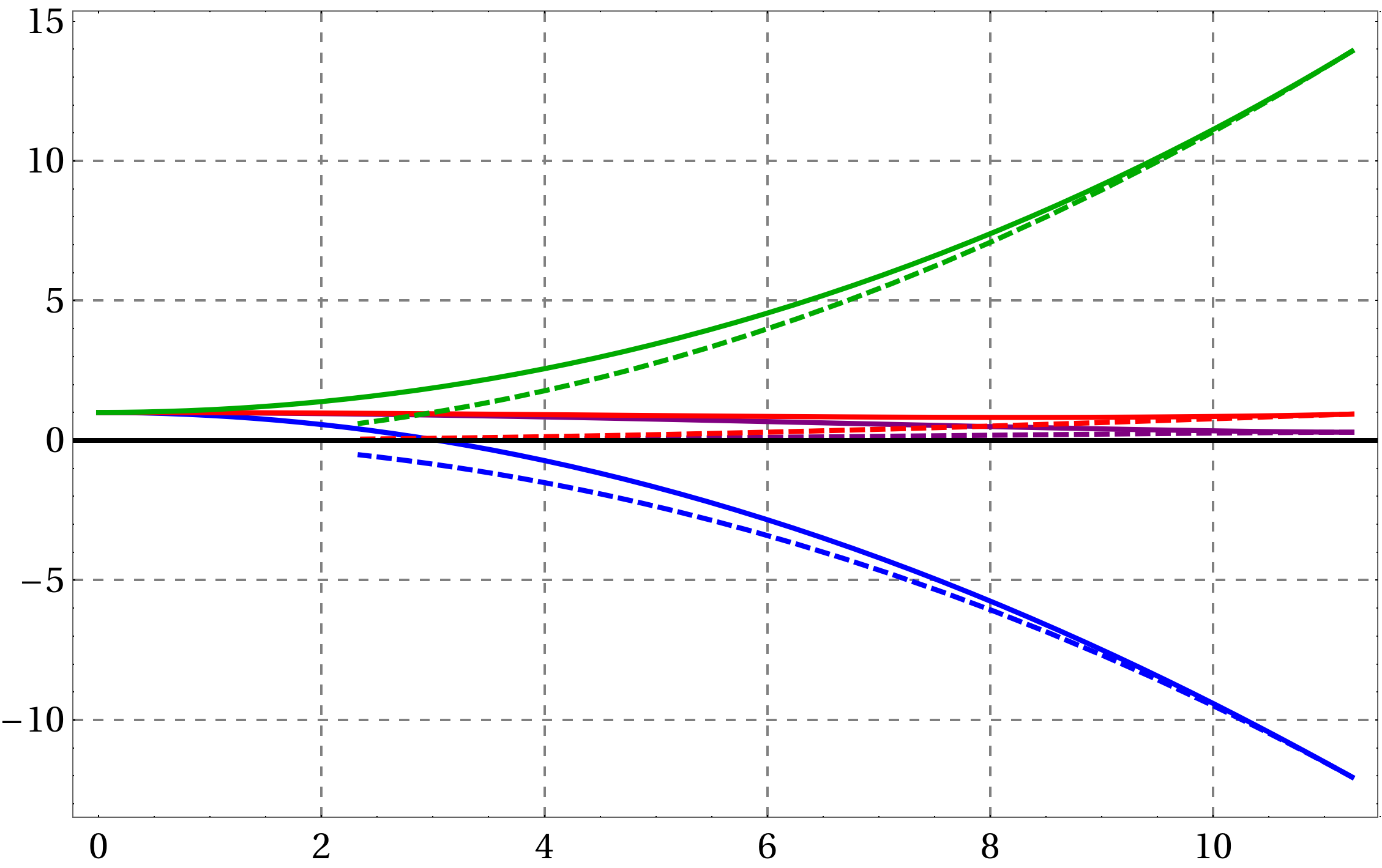}
 \put(-10,-10){\large $b/T^{2}$}
 \put(-370,220){\large $P^{\parallel}/P_{0}$}
\caption{Pressure $P^{\parallel}$ as a function of $b/T^{2}$, normalized with respect to its value $P_{0}$ at $b=0$ and $\varphi=0$. Each curve corresponds to a different renormalization scheme given by  $C_{sch}=-5$ (blue), $C_{sch}=-1/4$ (purple), $C_{sch}=0$ (red), and $C_{sch}=5$ (green).}
\label{PresurePar}
\end{figure}

\section{Thermodynamics}\label{thermo}

In this section we will compute and present the entropy density, specific heat, and free energy of the members of our family of solutions. We will address their implications in the last section.

\subsection{Entropy density}

\begin{figure}[ht!]
 \centering
 \includegraphics[width=0.8\textwidth]{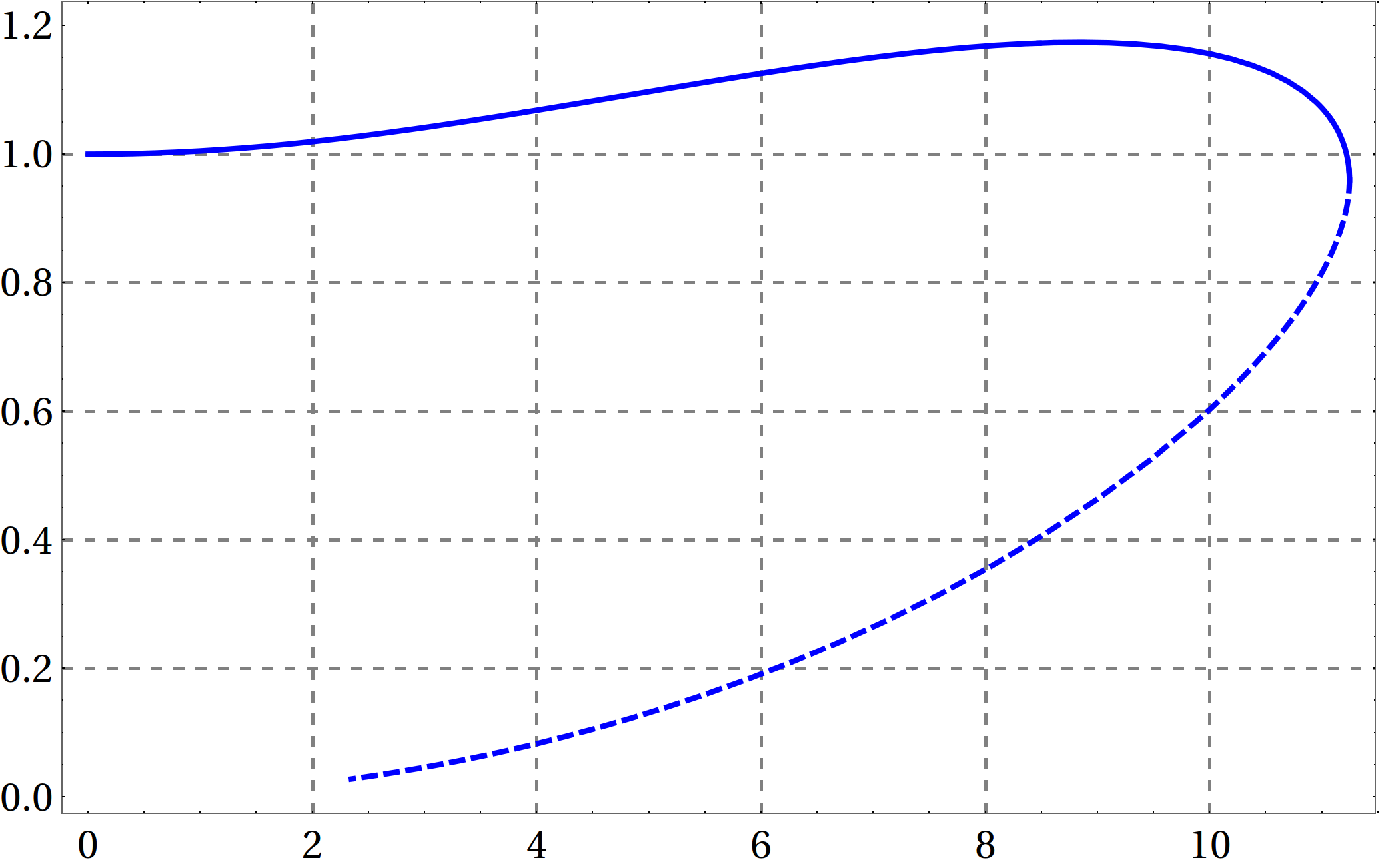}
 \put(-10,-10){\large $b/T^{2}$}
 \put(-370,220){\large $s/s_{0}$}
\caption{Entropy density as a function of $b/T^{2}$ normalized with respect to its value at $b=0$ and $\varphi=0$.}
\label{entropy1}
\end{figure}

\begin{figure}[ht!]
 \centering
 \includegraphics[width=0.8\textwidth]{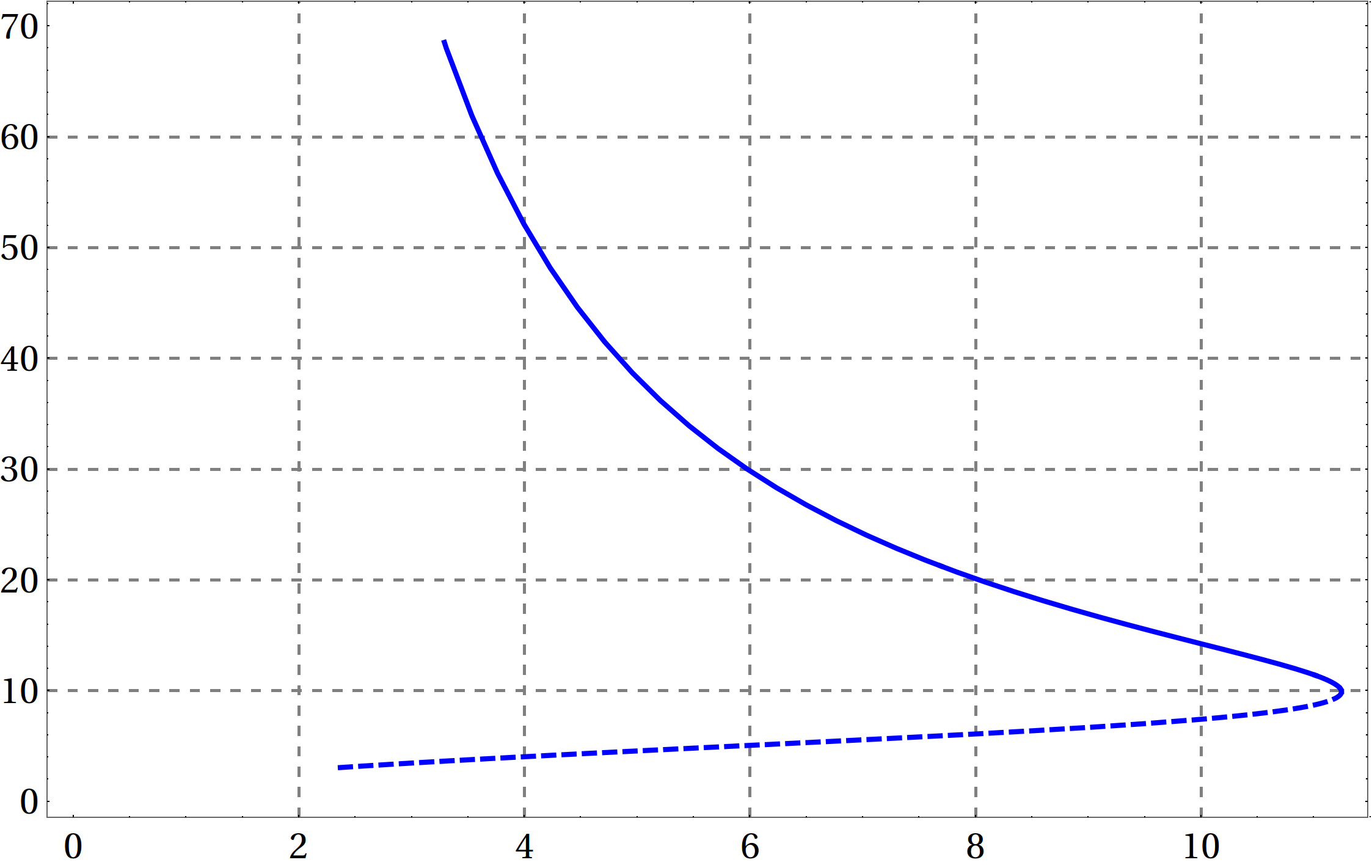}
  \put(-10,-10){\large $b/T^{2}$}
 \put(-370,220){\large $s/b^{3/2}$}
\caption{Dimensionless ratio $s/b^{3/2}$ as a function of $b/T^{2}$.}
\label{entropy2}
\end{figure}

The entropy density per unit of volume in the $(x,y,z)$ directions is given by the area of the horizon
\begin{equation}
s=\frac{A_{h}}{4G \text{vol}(x)}=\frac{N_{c}^{2}}{8\pi^{2}}\left(4\pi V(r_{h})\sqrt{W(r_{h})}\right),
\end{equation}
where $V(r_{h})$ and $W(r_{h})$ are the numerical metric functions evaluated at the horizon after they have been scale following subsection \ref{NumSol}, thus, the entropy density depends on the dimensionless ratio $b/T^{2}$, but, as it should be, is scheme independent. As a consistency check, we have verified that our numerical results do not depend independently on $b$ and $T$ if $b/T^{2}$ is kept fixed. Below we present the numerical results for the entropy density, where in Fig.(\ref{entropy1}) we normalized it with respect to its value at $b=0$ and $\varphi=0$
\begin{equation}
s_{0}=\frac{N_{c}^{2}}{2\pi}\left(\frac{9r_{h}^{2}}{4}\right)^{\frac{3}{2}}=\frac{\pi^{2}}{2}N_{c}^{2}T^{3},
\end{equation}
and plotted it as a function of $b/T^{2}$, whereas in Fig.(\ref{entropy2}) we show the dimensionless ratio $s/b^{3/2}$ as a function of $b/T^{2}$. The first of these figures is included to visualize the general behavior, while the second is directly related to the calculation of the specific heat in the next section.

\subsection{Specific heat}

\begin{figure}[ht!]
 \centering
 \includegraphics[width=0.8\textwidth]{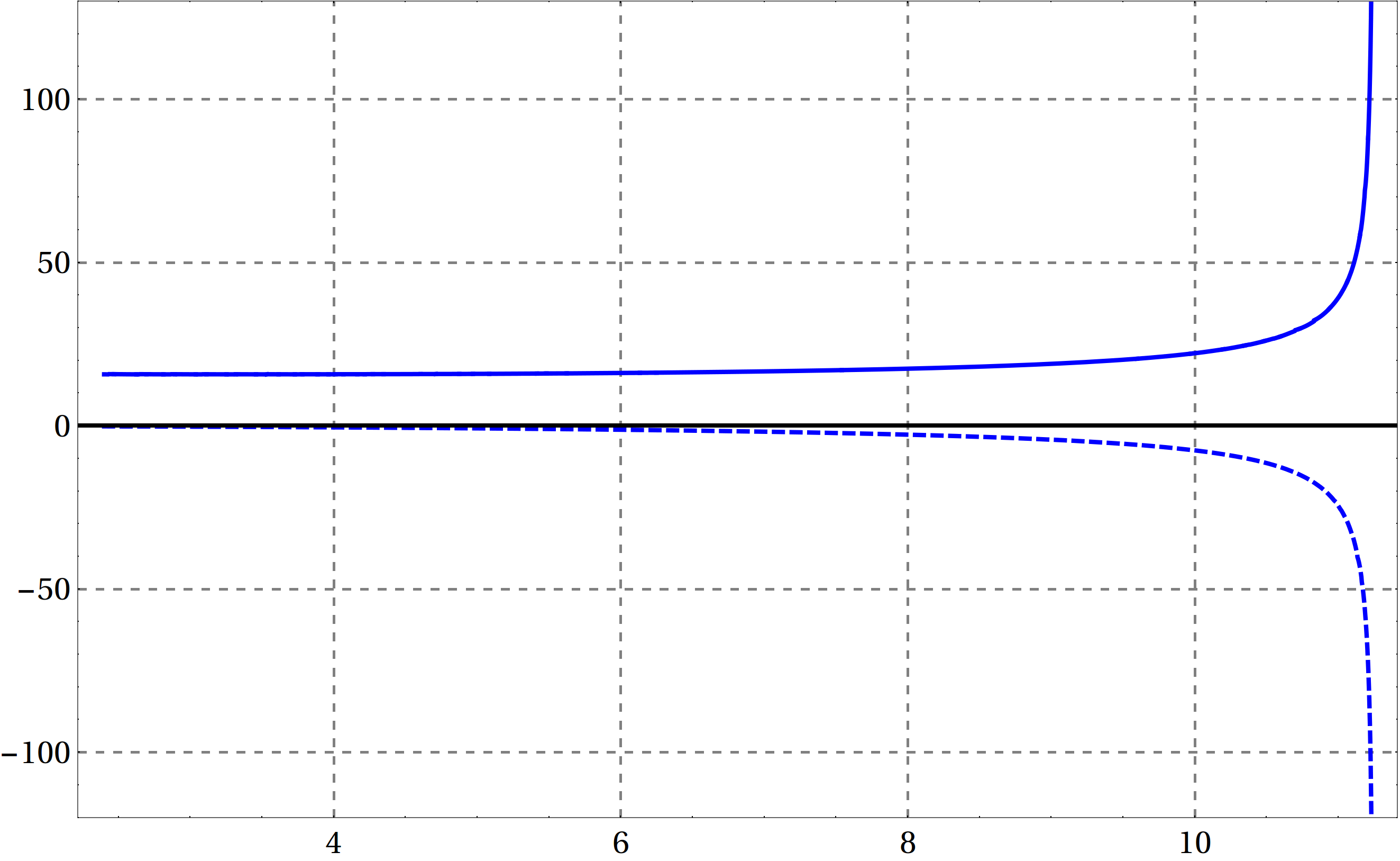}
  \put(-10,-10){\large $b/T{2}$}
 \put(-370,220){\large $C_{b}$}
\caption{Specific heat $C_{b}$ as a function of $b/T^{2}$.}
\label{CB}
\end{figure}

The specific heat $C_{b}$ at fixed magnetic field is given by
\begin{equation}
C_{b}=T\left(\frac{\partial s}{\partial T}\right)_{b}.
\end{equation}
To compute the derivative of the entropy density with respect to the temperature at fixed magnetic field, it is convenient to notice that the dimensionless ratio $s/b^{3/2}$ depends on $T$ and $b$ only through the dimensionless combination $b/T^{2}$, so that their relationship can be written as
\begin{equation}
\frac{s}{b^{3/2}}=H\left(\frac{b}{T^{2}}\right),
\end{equation}
for some function $H$, depicted in Fig.(\ref{entropy2}), that we can determine numerically, and which derivative with respect to its argument is related to the specific heat by
\begin{equation}
C_{b}=-\frac{2b^{5/2}}{T^{2}}H'\left(\frac{b}{T^{2}}\right).
\label{cbeq}
\end{equation}
We show the numerical result for $C_b$ as a function of $b/T^{2}$ in Fig.(\ref{CB}), where the scheme independence is inherited from the entropy density.

\subsection{Free Energy}

The free energy of the system is related to the fully renormalized on-shell action \eqref{ST} by
\begin{equation}
F=TS_{T}.
\label{Free}
\end{equation}
To evaluate the action in the family of solutions we found, it is convenient to separate it in a bulk integral
\begin{equation}
S_{bulk}=-\frac{1}{16\pi G_{5}}\int d^{5}x \sqrt{g}\left[R-\frac{1}{2}(\partial\varphi)^{2}+4\left(X^{2}+2X^{-1}\right)-X^{-2}(F)^{2}\right],
\label{S_bulk}
\end{equation}
and a boundary integral
\begin{equation}
\begin{split}
S_{bdry}=&\frac{1}{16\pi G_{5}}\int d^{4}x\sqrt{\gamma}\left(-2K+6+\varphi^{2}\left(1+\frac{1}{2\log{\epsilon}}\right)+F^{ij}F_{ij}\log{\epsilon}-C_{sch}F^{ij}F_{ij}\right)\\& +\frac{C_{sch}}{16\pi G_{5}}\int d^{4}x\sqrt{\gamma}\left(-F^{ij}F_{ij}+\frac{\varphi^{2}}{2\log^{2}{\epsilon}}\right).
\end{split}
\label{S_bdry}
\end{equation}

The infinite volume Vol$(x)$ resulting from the integration over the gauge theory directions can be factored out in both cases, and \eqref{S_bulk} can be further simplified using the equations of motion \eqref{Equations} to obtain
\begin{equation}
S_{bulk}=\frac{\text{Vol}(x)}{16\pi G_{5}T}\int_{r_{min}}^{r_{max}} dr \frac{4}{3}V\sqrt{W}\left(X^{-2}\frac{b^{2}}{V^{2}}+2(2X^{-1}+X^{2})\right),
\label{S_bulk_2}
\end{equation}
where $r_{min}$ is a radius close to the horizon and $r_{max}$ a cut-off radius near the boundary.

Concerning \eqref{S_bdry}, given that $r_{max}$ is eventually supposed to be send to the boundary, we can use the expansions \eqref{r_expansions} and only keep the leading terms
\begin{equation}
\begin{split}
S_{bdry}=&\frac{\text{Vol}(x)}{16\pi G_{5}T}(-2r_{max}^{4}-4U_{1}r_{max}^{3}-3U_{1}^{2}r_{max}^{2}-U_{1}^{3}r_{max}-\frac{4}{3}b^{2}\log(r_{max})\\& +2b^{2}\left(\frac{1}{3}-C_{sch}\right)-\frac{U_{1}^{4}}{8}-U_{4}+\mathcal{O}(r_{max}^{-1})).
\end{split}\label{Sbdry}
\end{equation}
In practice, subtracting \eqref{S_bulk_2} and \eqref{Sbdry} after evaluation involves dealing with the difference of two quantities that diverge as $r_{max}\rightarrow\infty$, increasing the numerical error considerably. To obtain an expression that is easily evaluated, it is convenient to replace the terms in \eqref{Sbdry} that are evaluated at $r_{max}$ by a radial integral from $r_{min}$ to $r_{max}$, plus this same terms evaluated at $r_{min}$, ending up with
\begin{equation}
\begin{split}
S_{bdry}=&-\frac{\text{Vol}(x)}{16\pi G_{5}T}\int_{r_{min}}^{r_{max}} dr\left(8r^{3}+12U_{1}r^{2}+6U_{1}^{2}r+U_{1}^{3}+\frac{4b^{2}}{3r}\right)\\&+\frac{\text{Vol}(x)}{16\pi G_{5}T}(-2r_{min}^{4}-4U_{1}r_{min}^{3}-3U_{1}^{2}r_{min}^{2}-U_{1}^{3}r_{min}-\frac{4}{3}b^{2}\log(r_{min})\\& +2b^{2}\left(\frac{1}{3}-C_{sch}\right)-\frac{U_{1}^{4}}{8}-U_{4}),
\end{split}
\label{S_bdry_2}
\end{equation}
where the contributions in \eqref{Sbdry} that do not depend on $r_{max}$ were left untouched.

\begin{figure}[ht!]
 \centering
 \includegraphics[width=0.8\textwidth]{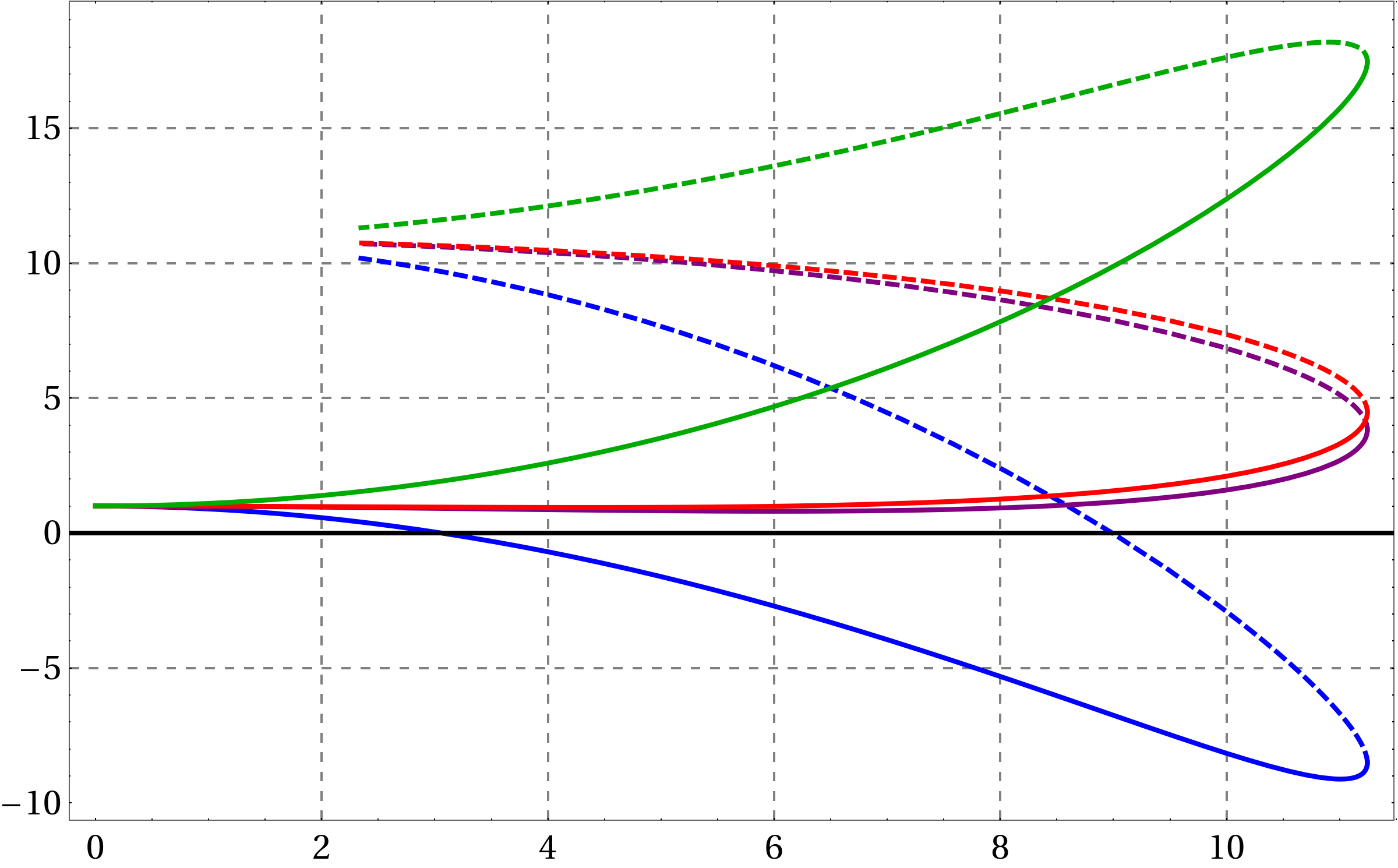}
 \put(-10,-10){\large $b/T^{2}$}
 \put(-370,220){\large $F/F_{0}$}
\caption{Free energy density $F$ as a function of $b/T^{2}$, normalized with respect to the value $F_{0}$ that it takes at vanishing magnetic field and scalar source. Each curve corresponds to a different renormalization scheme given by  $C_{sch}=-5$ (blue), $C_{sch}=-1/4$ (purple), $C_{sch}=0$ (red), and $C_{sch}=5$ (green).}
\label{FreeEnergy}
\end{figure}

Combining \eqref{S_bulk_2} and \eqref{S_bdry_2} with \eqref{Free} we obtain the final expression for the free energy density
\begin{equation}
\begin{split}
F=&\frac{N_{c}^{2}}{8\pi^{2}}\int_{r_{min}}^{r_{max}} dr \left[\frac{4}{3}V\sqrt{W}\left(X^{-2}\frac{b^{2}}{V^{2}}+2(2X^{-1}+X^{2})\right)-\left(8r^{3}+12U_{1}r^{2}+6U_{1}^{2}r+U_{1}^{3}+\frac{4b^{2}}{3r}\right)\right]\\&+\frac{N_{c}^{2}}{8\pi^{2}}\left(-2r_{min}^{4}-4U_{1}r_{min}^{3}-3U_{1}^{2}r_{min}^{2}-U_{1}^{3}r_{min}-\frac{4}{3}b^{2}\log(r_{min})+2b^{2}\left(\frac{1}{3}-C_{sch}\right)-\frac{U_{1}^{4}}{8}-U_{4}\right).
\end{split}
\end{equation}
In Fig.(\ref{FreeEnergy}) we present the results for the free energy normalized with respect to the value it takes at $b=0, \varphi=0$
\begin{equation}
F_{0}=-\frac{N_{c}^{2}\pi^{2}T^{4}}{8}.
\end{equation}

\section{Discussion}

As stated in the introduction, we were able to construct a family of backgrounds which members are dual to states in the gauge theory characterized by their temperature, the intensity of a background magnetic field, and the intensity with which an operator ${\cal{O}}_{\varphi}$ of scaling dimension $\Delta=2$ is sourced.

The first thing we found is that for any given temperature and value of the source of ${\cal{O}}_{\varphi}$, there exists a maximum intensity $b_c$ for the magnetic field, above which the state is unstable. The analysis we performed to determine the existence of this $b_c$ is in essence the same as the one done in 
\cite{Horowitz:2016ezu} about cosmic censorship in 4D, where the solutions to Einstein-Maxwell with AdS boundary conditions develop a naked singularity for intensities of an electric field higher than a certain value. Interestingly, in \cite{Crisford:2018qkz} the same authors develop a vacuum analog of \cite{Horowitz:2016ezu}, where they consider a differential rotation on the boundary metric. Keeping the profile of the differential rotation fixed, but increasing the overall amplitude, it is shown that smooth solutions only exist up to a finite maximum amplitude. The reason why we find this of particular relevance is that in the 10D uplift \cite{Avila:2019pua} of our family of solutions the magnetic field is encoded as an infinitesimal rotation in the compact part of the 10D spacetime, so, even if in \cite{Crisford:2018qkz} the rotation is in the non compact directions, the findings seem equivalent.


There are other calculations that indicate that the existence of a maximum for the intensity of the magnetic field is not uncommon. For instance, even if not thought of as a consistent truncation, in \cite{Rougemont:2015oea} the authors build a 5D model that includes a constant magnetic field and a scalar field, which they refer to as a dilaton. In order to obtain agreement with lattice QCD data, a potential is chosen for the scalar such that its scaling dimension is $\Delta=3$. In this context it is also found that for a given value of the dilaton at the horizon, there exists a maximum intensity for the magnetic field that allows the solution to asymptoticaly approach AdS.

Working at fixed temperature, we find that for any intensity $b$ below our $b_c$ there are two possible states that differ in the value of the vacuum expectation value of ${\cal{O}}$, and through our thermodynamic analysis we were able to prove that one of them is favored over the other. We base this claim on the fact that, independently of the renormalization scheme, we consistently found that one of the branches has lower entropy, higher free energy, and negative specific heat, indicative of a thermal instability. Furthermore, by using expression \eqref{cbeq}, that is scheme independent, and noticing that the curve in Fig.(\ref{entropy2}) becomes vertical at $b/T^{2}=b_{c}/T^{2}$, we see that the specific heat becomes infinite at $b_c$, indicating the presence of a phase transition. We show this, up to the numerical precision we achieved, in Fig.(\ref{CB}). This resembles the results in 
\cite{Banks:2015aca,Banks:2016fab}, where a scalar field dual to an operator of scaling dimension $\Delta=2$ is added to the anisotropic background \cite{Mateos:2011ix,Mateos:2011tv}. Following the same motivation as us, they turn off the source of the operator and hence reduce the dimensionless parameters to only $a/T$, similar to our $b/T^{2}$, and are able to write $\langle \mathcal{O}_{\varphi}\rangle$ as a function of $a/T$. They describe how it is possible to turn off the scalar field continuously while keeping a non-vanishing anisotropy, and demonstrate that there is a critical $a/T$ above which it is thermodynamically preferable to keep the scalar on. They also claim that even if they have not shown that their solution is a consistent truncation from a ten dimensional theory, it is straightforward to do so. It is relevant to point out that the phase to which our system transitions for fields stronger than $b_c$ is not the unstable branch that we find, but a different state, dual to the time dependent background discussed in section \ref{PhysParam} that is still to be determined in future work, where our current findings on the free energy will play an important role.

From the stress-energy tensor that we computed we saw that there is a conformal anomaly, which is of particular interest in views of the proposal in \cite{Basar:2012bp}, where it is claimed that an anomaly in the presence of a strong magnetic field can explain the excess of photons reported in the ALICE experiment \cite{Wilde:2012wc}. To explore this speculation we are currently computing the spectrum of direct photon production in such a scenario to determine if there is an enhancement and how does it compare with the one presented in \cite{Arciniega:2013dqa}. Within the context of our own work, it is the existence of this anomaly that makes it not surprising to find that some of the physical quantities that we computed turned out to be dependent on the renormalization scheme, that was encoded in the magnitude of the finite term that we could add to the finite Euclidean action.

\section{Acknowledgments}

It is a pleasure for us to thank Gary Horowitz for a very helpful discussion about the present work during his visit to UNAM, and for pointing out references \cite{Crisford:2017gsb} and \cite{Crisford:2018qkz} which inspired our analysis of the unstable behavior. It is also a pleasure to thank Alberto G\"uijosa, for helping with our physical understanding of the system and the suggestion to modify the title of this work, and Francisco Nettel for a careful revision of this manuscript. We also acknowledge partial financial support from PAPIIT IN113618, UNAM.

\appendix

\section{Near boundary analysis}
 
\subsection{Holographic renormalization}
\label{AppA}

As it is very commonly the case on the AdS/CFT correspondence, the direct on-shell evaluation of the Euclidean action \eqref{euclidean-action} diverges, so to have a well defined variational principle, we need to renormalize it through a method that has been extensively studied \cite{Skenderis:2002wp}. The first step is to analyze the behavior of the solution near the boundary, which is more conveniently done in the Fefferman-Graham coordinate where the metric takes the form
\begin{equation}
ds_{5}^{2}=\frac{du^{2}}{u^{2}}+\gamma_{ij}(u)dx^{i}dx^{j}=\frac{1}{u^{2}}\left(du^{2}+g_{ij}(u)dx^{i}dx^{j}\right),
\label{metric_5_FG}
\end{equation}
encoding the geometric information in $g_{ij}$.

We solve the equations of motion \eqref{EOM_fondoa}, \eqref{EOM_fondob}, and \eqref{EOM_F}, by a power series method around $u=0$, obtaining the expansions for $g, \varphi,$ and $F$ given by
\begin{eqnarray}
&& g_{ij}(u)={g_{ij}}_{(0)}+({g_{ij}}_{(4)}+h_{ij}\log{u}+H_{ij}\log^{2}{u})u^{4}+\mathcal{O}(u^{6}),
\cr
&& \varphi(u)=u^{2}(\varphi_{(0)}+\psi_{(0)}\log{u}+(\varphi_{(2)}+\psi_{(2)}\log{u}+ \Psi_{(2)}\log^{2}{u})u^{2})+\mathcal{O}(u^{6})
\cr
&& F_{u\nu}=0, \qquad F_{ij}=F_{ij}(t,x,y,z),
\label{Fondo_FG}
\end{eqnarray}
where
\begin{eqnarray}
&& \Psi_{(2)}=-\frac{\psi_{(0)}^{2}}{2\sqrt{6}},
\cr
&& \psi_{(2)}=\frac{\psi_{(0)}}{\sqrt{6}}(\psi_{(0)}-\varphi_{(0)}),
\cr
&& \varphi_{(2)}=-\frac{1}{\sqrt{6}}\left(\frac{1}{2}F_{ik}F_{jl}{g^{ij}}_{(0)}{g^{kl}}_{(0)}+\frac{1}{2}\varphi_{(0)}^{2}+\frac{3}{4}\psi_{(0)}^{2}-\varphi_{(0)}\psi_{(0)}\right),
\cr
&& {g^{ij}}_{(0)}{g_{ij}}_{(4)}=\frac{1}{12}F_{ik}F_{jl}{g^{ij}}_{(0)}{g^{kl}}_{(0)}-\frac{1}{3}\varphi_{(0)}^{2}-\frac{1}{24}\psi_{(0)}^{2},
\cr
&& h_{ij}=\frac{1}{4}{g_{ij}}_{(0)}F_{nk}F_{ml}{g^{nm}}_{(0)}{g^{kl}}_{(0)}-F_{ik}F_{jl}{g^{kl}}_{(0)}-\frac{1}{6}{g_{ij}}_{(0)}\varphi_{(0)}\psi_{(0)},
\cr
&& H_{ij}=-\frac{1}{12}{g_{ij}}_{(0)}\psi_{(0)}^{2},
\label{solucion_FG}
\end{eqnarray}
and any non listed coefficient up to the specified order is equal to zero. We notice in passing that, similarly to \cite{Janiszewski:2015ura}, the first two terms in the expression for $h_{ij}$ constitute the stress-energy tensor of the electromagnetic field in the boundary theory.

Given the specificities of our case, we have taken $F$ to only depend on the gauge theory directions, even if it is worth noticing that $F\wedge F=0$ imposes constrains on this field that will not play a role in the following calculations, but that anyway are satisfied by the constant magnetic field solutions studied here. Also note that the leading order in the expansion for $\varphi$ is $u^{2}$, which means that the field saturates the BF bound \cite{Breitenlohner:1982jf} with $m^{2}=-4$, and thus it is dual to a boundary operator ${\cal{O}}$ of scaling dimension $\Delta=2$. For this kind of field, $\psi_{(0)}$ in \eqref{Fondo_FG} is dual to the source of ${\cal{O}}$, while $\varphi_{(0)}$ is dual to its vacuum expectation value \cite{Bianchi:2001kw}.  

Since we are interested in isolating the divergences of the on-shell Euclidean action as $u\rightarrow 0$, we substitute \eqref{Fondo_FG} into \eqref{euclidean-action} and integrate from a radial cut-off $\epsilon$ to an arbitrary $u_{\text{max}}$, which even if bigger than $\epsilon$, is still close to the boundary and remains fixed. As $\epsilon\rightarrow 0$, the approximated integral just described diverges exactly as the full on-shell Euclidean action does, behavior that is then captured by the diverging terms in the boundary integral at $u=\epsilon$ given by
\begin{equation}
S_{\epsilon}=-\frac{1}{16\pi G_{5}}\int d^{4}x\sqrt{g_{(0)}}\left(\frac{1}{\epsilon^{4}}a_{(0)}+a_{(1)}\log{\epsilon}+a_{(2)}\log^{2}{\epsilon}+\mathcal{O}(\epsilon^{0})\right),
\label{div_action}
\end{equation}
where
\begin{eqnarray}
&& a_{(0)}=6,
\cr
&& a_{(1)}=3{g^{ij}}_{(0)}h_{ij}+F_{ik}F_{jl}{g^{ij}}_{(0)}{g^{kl}}_{(0)}+\frac{1}{2}\psi_{(0)}(4\varphi_{(0)}+\psi_{(0)}),
\cr
&& a_{(2)}=3{g^{ij}}_{(0)}H_{ij}+\psi_{(0)}^{2},
\end{eqnarray}
which further simplify to
\begin{eqnarray}
&& a_{(0)}=6,
\cr
&& a_{(1)}=F_{ik}F_{jl}{g^{ij}}_{(0)}{g^{kl}}_{(0)}+\frac{1}{2}\psi_{(0)}^{2},
\cr
&& a_{(2)}=0,
\label{div}
\end{eqnarray}
when the solution \eqref{solucion_FG} is used.

The next step is to invert the series \eqref{Fondo_FG} to express the coefficients involved in \eqref{div} in terms of the fields $g_{ij}(u), \varphi(u),$ and $F$, and its derivatives. To the relevant order for this calculation, and taking one step further to use $\gamma_{ij}(u)$ instead of $g_{ij}(u)$, we get 
\begin{equation}
{g_{ij}}_{(0)}=\epsilon^{2}\gamma_{ij}, \qquad {g^{ij}}_{(0)}=\frac{\gamma^{ij}}{\epsilon^{2}}, \qquad \psi_{(0)}=\frac{\varphi}{\epsilon^{2}\log{\epsilon}}, \qquad \sqrt{g_{(0)}}=\epsilon^{4}\sqrt{\gamma}\left(1+\frac{1}{6}\varphi^{2}\right),
\end{equation}
reducing the expressions in \eqref{div} to the final form
\begin{equation}
a_{(0)}=6, \qquad a_{(1)}=\frac{1}{\epsilon^{4}}\left(F^{ij}F_{ij}+\frac{1}{2\log^{2}{\epsilon}}\varphi^{2}\right), \qquad a_{(2)}=0,
\label{coefficients}
\end{equation}
where the indexes are raised and lowered using the boundary metric $\gamma_{ij}$. By substituting \eqref{coefficients} in \eqref{div_action} and discarding finite terms we obtain
\begin{equation}
S_{\epsilon}=-\frac{1}{16\pi G_{5}}\int d^{4}x\sqrt{\gamma}\left(6+\varphi^{2}\left(1+\frac{1}{2\log{\epsilon}}\right)+F^{ij}F_{ij}\log{\epsilon}\right),
\end{equation}
which is the negative of the counterterm \eqref{counters} used in section \ref{STS}.

\subsection{Stress-energy tensor and scalar condensate}
\label{AppB}

With the renormalized action on hand, it is possible to obtain a number of observables in the gauge theory. In this subsection of the appendix we specifically compute the vacuum expectation value $\langle \mathcal{O}_{\varphi}\rangle$ of the operator dual to the scalar field, and the expectation value $\langle T_{ij}\rangle$ of the stress-energy tensor in the states dual to the members of the family of solutions that we found.

The holographic dictionary states that $\langle \mathcal{O}_{\varphi}\rangle$ is given by \cite{Skenderis:2002wp,Bianchi:2001kw}
\begin{equation}
\langle \mathcal{O}_{\varphi}\rangle=\lim_{\epsilon\rightarrow 0}\left(\frac{\log{\epsilon}}{\epsilon^{2}}\frac{1}{\sqrt{\gamma}}\frac{\delta S}{\delta\varphi}\right),
\end{equation}
where the presence of the logarithmic term is due to the scalar saturating the BF bound. Taking the variation of the total action \eqref{ST} with respect to the scalar field we obtain
\begin{equation}
16\pi G_{5}\langle \mathcal{O}_{\varphi}\rangle=\lim_{\epsilon\rightarrow 0}\frac{\log{\epsilon}}{\epsilon^{2}}\left(-u\partial_{u}\varphi+\varphi\left(2+\frac{1}{\log{\epsilon}}+\frac{C_{sch}}{\log^{2}{\epsilon}}\right)\right),
\end{equation}
which, with the help of the asymptotic expansions \eqref{Fondo_FG}, rewrites as
\begin{equation}
16\pi G_{5}\langle \mathcal{O}_{\varphi}\rangle=\varphi_{(0)}+C_{sch}\psi_{(0)},
\label{condensate}
\end{equation}
showing that the VEV of $\mathcal{O}_\varphi$ is given by $\varphi_{(0)}$ while, as usual, the $\psi_{(0)}$ contribution is scheme dependent.

The holographic dictionary relates the stress-energy tensor in the gauge theory to variations of the action with respect to the boundary metric, namely \cite{Skenderis:2002wp,Bianchi:2001kw}
\begin{equation}
\langle T_{ij}\rangle=\lim_{\epsilon\rightarrow 0}\left(\frac{1}{\epsilon^{2}}\frac{2}{\sqrt{\gamma}}\frac{\delta S}{\delta\gamma^{ij}}\right).\label{tij}
\end{equation}

Taking $S$ in \eqref{tij} to be the total action \eqref{ST}, and doing some algebra, we obtain
\begin{equation}
\begin{split}
16\pi G_{5}\langle T_{ij}\rangle=&\lim_{\epsilon\rightarrow 0}\frac{2}{\epsilon^{2}}(-K_{ij}+2F_{ik}{F_{j}}^{k}(\log{\epsilon}-C_{sch})\\&-\frac{1}{2}\gamma_{ij}(-2K+6+\varphi^{2}(1+\frac{1}{2\log{\epsilon}}+\frac{C_{sch}}{2\log^{2}{\epsilon}})+F^{kl}F_{kl}(\log{\epsilon}-C_{sch}))),
\end{split}
\end{equation}
which after using the asymptotic expansions \eqref{Fondo_FG} rewrites as
\begin{equation}
\begin{split}
16\pi G_{5}\langle T_{ij}\rangle=&4{g_{ij}}_{(4)}+h_{ij}(1+4C_{sch})+6C_{sch}H_{ij}\\&-{g_{ij}}_{(0)}\left({g^{kl}}_{(0)}(4{g_{kl}}_{(4)}+h_{kl})+\varphi_{(0)}(\varphi_{(0)}+\psi_{(0)}(1-\frac{2}{3}C_{sch}))\right),
\end{split}
\label{stressFG}
\end{equation}
that is our final expression for the scheme dependent expectation value of the stress-energy tensor. We see here that as in \cite{Janiszewski:2015ura}, choosing $C_{sch}=-1/4$ would eliminate the explicit contribution of the electromagnetic field to the boundary stress-energy tensor encoded in $h_{ij}$. It is worth noticing that given (\ref{solucion_FG}), the terms involving ${g_{ij}}_{(4)}$ in (\ref{stressFG}) also contain a contribution proportional to the electromagnetic part of the stress-energy tensor, introducing subtleties concerning how much of it is proper to the plasma itself.

The trace of the stress-energy tensor is given by contracting \eqref{stressFG} with ${g_{(0)}}^{ij}$ and results in
\begin{equation}
16\pi G_{5}\langle {T^{i}}_{i}\rangle=-F_{ik}F_{jl}{g^{ij}}_{(0)}{g^{kl}}_{(0)}-\psi_{(0)}\left(2\varphi_{(0)}+\frac{1}{2}\psi_{(0)}(4C_{sch}-1)\right),
\label{traceFG}
\end{equation}
which is non-zero, showing the existence of a conformal anomaly in the theory.

\subsection{Scaling}
\label{AppA3}

Here we will determine how $\langle T_{ij}\rangle$ and $\langle \mathcal{O}_{\varphi}\rangle$ transform under a scaling of the form
\begin{equation}
x^{i}\rightarrow kx^{i}, \qquad u\rightarrow ku,
\label{scaletransformation}
\end{equation}
which in terms of the physical parameters is equivalent to
\begin{equation}
b\rightarrow k^{2}b, \qquad T\rightarrow kT,
\end{equation}
for $k$ a positive real number.

Following \cite{Bianchi:2001de}, we note that the FG form of the metric \eqref{metric_5_FG} is preserved under \eqref{scaletransformation} as long as the coefficients appearing in \eqref{Fondo_FG} transform as
\begin{eqnarray}
&& {g_{ij}}_{(4)}\rightarrow k^{4}({g_{ij}}_{(4)}+h_{ij}\log{k}+H_{ij}\log^{2}{k}), 
\cr
&& h_{ij}\rightarrow k^{4}(h_{ij}+2H_{ij}\log{k}),
\cr
&& H_{ij}\rightarrow k^{4}H_{ij},\label{coeffsca}
\\
&& \varphi_{(0)}\rightarrow k^{2}(\varphi_{(0)}+\psi_{(0)}\log{k}),
\cr
&& \psi_{(0)}\rightarrow k^{2}\psi_{(0)}.\nonumber
\end{eqnarray}

Using \eqref{coeffsca} in \eqref{condensate} and \eqref{stressFG} we respectively get that the transformation rule for $\langle \mathcal{O}_{\varphi}\rangle$ is given by
\begin{equation}
\langle \mathcal{O}_{\varphi}\rangle\rightarrow k^{2}\langle \mathcal{O}_{\varphi}\rangle +\psi_{(0)}k^{2}\log{k},
\end{equation}
and the one for $\langle T_{ij}\rangle$ by
\begin{eqnarray}
\langle T_{ij}\rangle\rightarrow & k^{4}\langle T_{ij}\rangle +k^{4}\log^{2}{k}\left[4H_{ij}-{g_{ij}}_{(0)}\left(4{g^{kl}}_{(0)}H_{kl}+\psi_{(0)}^{2}\right)\right] \cr & +k^{4}\log{k}[4h_{ij}+2H_{ij}(1+4C_{sch})\cr &-{g_{ij}}_{(0)}\left({g^{kl}}_{(0)}(4h_{kl}+2H_{kl})+\psi_{(0)}(\psi_{(0)}(1-\frac{2}{3}C_{sch})+2\varphi_{(0)})\right)].
\end{eqnarray}
These rather complicated transformation rules simplify enormously when the source $\psi_{(0)}$ for $\langle \mathcal{O}_{\varphi}\rangle$ is turn off, in which case they reduce to
\begin{eqnarray}
&& \langle T_{ij}\rangle\rightarrow k^{4}\langle T_{ij}\rangle +4k^{4}h_{ij}\log{k},
\cr
&& \langle \mathcal{O}_{\varphi}\rangle\rightarrow k^{2}\langle \mathcal{O}_{\varphi}\rangle .
\end{eqnarray}

\end{document}